\documentclass{article}
\usepackage{amsmath,amsthm,amssymb,array,latexsym,cite,epsfig,final}

\numberwithin{equation}{section}

\input eymA.def

\begin{document}

\title{On all possible static spherically symmetric EYM\\
solitons and black holes
\thanks{Supported in part by NSERC grant A8059.}
\thanks{PACS: 04.40.Nr, 11.15.Kc}
\thanks{2000 \emph{Mathematics Subject Classification} 
Primary 83C20, 83C22; Secondary 53C30, 17B81.}
}
\author{
Todd A. Oliynyk \thanks{toliynyk@ualberta.ca}\\[3mm]
%\and
H.P. K\"{u}nzle \thanks{hp.kunzle@ualberta.ca}\\[3mm]
Department of Mathematical Sciences, University of Alberta\\
Edmonton, Canada T6G 2G1}

\date{}
\maketitle
\begin{abstract}
  We prove local existence and uniqueness of static spherically symmetric
  solutions of the Einstein-Yang-Mills equations for any action of the
  rotation group (or $SU(2)$) by automorphisms of a principal bundle over
  space-time whose structure group is a compact semisimple Lie group $G$.
  These actions are characterized by a vector in the Cartan subalgebra of $\g$
  and are called regular if the vector lies in the interior of a Weyl chamber.
  In the irregular cases (the majority for larger gauge groups) the boundary
  value problem that results for possible asymptotically flat soliton or black
  hole solutions is more complicated than in the previously discussed regular
  cases. In particular, there is no longer a gauge choice possible in general
  so that the Yang-Mills potential can be given by just real-valued functions.
  We prove the local existence of regular solutions near the singularities of
  the system at the center, the black hole horizon, and at infinity, establish
  the parameters that characterize these local solutions, and discuss the set
  of possible actions and the numerical methods necessary to search for global
  solutions. That some special global solutions exist is easily derived from
  the fact that $\su(2)$ is a subalgebra of any compact semisimple Lie
  algebra. But the set of less trivial global solutions remains to be
  explored.
\end{abstract}

\sect{intro}{Introduction}

The classical interaction between gravitational and \YM\ fields is
described by a complicated highly nonlinear field equations which,
even when reduced to a system of ordinary differential equations in
the static spherically symmetric case, leads to many interesting
mathematical and physical problems.  Physically these solutions have
shown that equilibrium configurations of black holes can be much more
complicated than had previously been thought since mass, charge and
angular momentum are clearly not enough to characterize them.  There
is even numerical evidence for the existence of non spherical
axisymmetric static black holes \cite{k6160}. Mathematically, an
analysis of the solution space of the static spherically symmetric
equations requires an interesting combination of geometrical,
algebraic, analytic and numerical techniques. The global solutions of
$SU(2)$-EYM equations have been extensively studied analytically
\cite{k5201,k5061,k6135,k6499}. For a fairly comprehensive summary of the
substantial literature on the subject we refer to the review article
\cite{k6373}. Almost all these investigations, however, have only
studied the gauge groups $SU(2)$ and occasionally $SU(n)$ for $n>2$,
and only for the most obvious ansatz for a spherically symmetric gauge
field.

But spherical symmetry for \YM\ fields is more complicated to define
than for tensor fields on a manifold because there is no unique way to
lift an isometry on space-time to the bundle space.  The only natural
way to define spherical symmetry of a \YM\ field is to require that it
be invariant under an action of the rotation group by automorphisms on
the principal bundle whose structure group is the gauge group $G$.  A
conjugacy class of such automorphisms is characterized by a generator
$\Lo$ which is an element of a Cartan subalgebra $\h$ of the
complexified Lie algebra $\g$ of $G$ \cite{k5109,k5281}. If one
restricts consideration to fields which are bounded at the center or,
in the presence of black hole fields, to those for which the
\YM-curvature falls off sufficiently fast at infinity then $\Lo$ must
be a \emph{defining vector} of an $\sL(2)$-subalgebra of $\g$.

One of these classes of actions of the symmetry group is somewhat
distinguished. It corresponds to a principal defining vector in Dynkin's
terminology and we will call it a \emph{principal action}. Almost all
work for larger gauge groups has been done for this case
\cite{hka28,k5749, k6485,k6299}. A slightly bigger class of actions,
which we call \emph{regular}, consists of those for which the defining
vector lies in the interior of a Weyl chamber. For those, for example,
Brodbeck and Straumann \cite{k4295,k5626} proved that all bounded
static asymptotically flat solutions are unstable against time
dependent perturbations.

Of course, one is interested in global solutions of the boundary value problem
that results from demanding boundedness at the singularities of the
differential equations at the center or the horizon and at infinity. This
global existence has long been established for $G=SU(2)$ by Smoller et al.
\cite{k5061,k5201} (see also \cite{k5428}). It is easy to ``imbed'' these
solutions into the set of solutions of the theory with arbitrary compact
semi-simple $G$ since the latter always has $SU(2)$ subgroups. The problem
is thus not to prove that global solutions exist but to explore the global
solution space and, hopefully, characterize different types of solutions, for
example, by their behavior near the center or near infinity.

In \cite{eymg} we have therefore considered and solved the local existence
problem for bounded solutions near the singularities for the \emph{regular}
symmetry group actions. We have identified the set of ``initial conditions''
that must be given at $r=0$ or $r=\infty$ to guarantee the local existence and
uniqueness of a bounded solution. For arbitrary gauge groups this required an
fairly intricate application of the $\Sl{2}$ representation theory on the
(complexified) Lie algebra $\g$ of $G$.

The purpose of the present paper is simply to extend these results to the case
where the symmetry group action need not be regular. It turns out that the
situation is qualitatively not very different. There are still similar
algebraic equations that restrict the possible initial data. But the number of
functions that will characterize the gauge potential is no longer just the
rank of the Lie algebra $\g$ or of one of its subalgebras. Even the set of
reduced field equations can no longer be determined simply from the structure
of the Cartan subalgebra (one needs access to all the Lie brackets of $\g$).
Moreover, a simple gauge choice that allows one to describe the potential in
terms of rank($\g$) real functions is no longer available. Complex functions
must be allowed which increases substantially the number of parameters to be
determined in a numerical solution of the boundary value problem.

In section \ref{gpotential} we review definitions and previous results
mostly from \cite{eymg} and then describe in section \ref{cmpt} how
the field equations can be handled computationally. In section
\ref{class} we give some examples of the possible symmetry group
actions for low dimensional gauge groups. The methods of section
\ref{cmpt} do not lend themselves easily to a general existence
proof. This needs to be done differently. In section \ref{alg} we
state and prove some algebraic lemmas that needed in section \ref{lep}
for the local existence and uniqueness proofs.  We conclude by giving
a preliminary example of a numerical irregular solution showing that
imaginary parts of the functions $w_{\alpha}(r)$ develop even if the
values $w_{\alpha}(0)$ or $w_{\alpha}(\infty)$ are all real. We have
not yet found a (nontrivial) global asymptotically flat numerical
solution.

\sect{gpotential}{\YM\ potentials and field equations}

Let $P$ be a principal bundle with a compact semi-simple structure group $G$
over a static spherically symmetric space-time manifold. For simplicity we
consider only actions of the group $SU(2)$ by principal bundle automorphisms
on $P$ that project onto the action of $SO(3)$ on space-time which defines the
spherical symmetry\footnote{This ignores some interesting effects due to the
  fact that $SO(3)$ is not simply connected. For an analysis of $SO(3)$
  actions on $SU(2)$ bundles see \cite{k6217}.}.  
  
Equivalence classes of these spherically symmetric $G$-bundles are in
one-to-one correspondence to conjugacy classes of homomorphisms of the
isotropy subgroup, $U(1)$ in this case, into $G$. The latter, in turn, are
given by their generator $\La_{3}$, the image of the basis vector
$\tau_{3}$ of $\su(2)$ (where $\{\tau_{i}\}$ is a standard basis with
$[\tau_{i},\tau_{j}]=\epsilon^{k}_{\phantom{k}ij}\tau_{k}$), lying in an
integral lattice $I$ of a Cartan subalgebra $\h$ of the Lie algebra $\g_{0}$
of $G$. This vector $\La_{3}$, when nontrivial, then characterizes up to
conjugacy an $\su(2)$ subalgebra. (See, for example, \cite{k4293}; we
follow the notation of this text and also of \cite{k5157}.)

It is convenient to pass to the complexified Lie algebra $\g$ of $\g_{0}$ and
define 
\leqn{lam0}{ \Lo := 2i\La_{3}\;. }
We now regard $\g_{0}$ as a compact real form of $\g$ which defines the
conjugation $c$ on $\g$ (a Lie algebra automorphism satisfying $c\circ
c=\id$). Then we can write $\g=\g_{0}+i\g_{0}$, $c(X+iY)=X-iY$ if
$X,Y\in\g_{0}$, and also $\h=\h_{0}+i \h_{0}$ where $\h_{0}$ is a Cartan
subalgebra of $\g_{0}$. Moreover,
\leqn{conj}{
 \Lo \in i\h_{0}, \quad c(\Lo) = -\Lo \;.
}

We will use the following notation from Lie algebra theory (following
\cite{k5157},\cite{k6494},\cite{k4293}): $\Ad_{g}:\g_{0}\ra\g_{0}
\;\forall\;g\in G$ is the adjoint action of the Lie group $G$ on its
(real) Lie algebra $\g_{0}$ while $\ad:\g\ra\gl(\g)$ is defined by
$\ad(X)(Y):=[X,Y]$. Define also the centralizer of $X$ in $\g$ by
\leqn{centr}{ \g^{X}:=\{Y\in\g\;|\;[X,Y]=0\} } and write $\g_{0}^{X}$
for the corresponding centralizer of the real Lie algebra.

Wang's theorem \cite{k4872,k0929} on connections that are invariant
under actions transitive on the base manifold has been adapted to spherically
symmetric space-time manifolds by Brodbeck and Straumann \cite{k5281}. They
show that in a Schwarzschild type coordinate system $(t,r,\theta,\phi)$ and
the metric
\leqn{metric}{g = -N S^{2} dt^{2}+
  N^{-1}dr^{2}+r^{2}(d\theta^{2}+\sin^{2}\theta d\phi^{2})}
a gauge can always be chosen such that the $\g_{0}$-valued \YM-connection form
is locally given by $A=\tilde{A}+\hat{A}$ where 
\leqn{Atilde}{
\tilde{A} = N(t,r)S(t,r)\Ac(t,r) dt + \Bc(t,r) dr
}
is an $\Ad(\La_{3})$-invariant 1-form (i.e. with values in $\g_{0}^{\La_{3}}$)
on the quotient space parametrized by the $r$ and $t$ coordinates and
\leqn{Ahat}{
  \hat{A} = \La_{1} d\theta + (\La_{2}\sin\theta + \La_{3}\cos\theta)d\phi\;.
} 
Here $\La_{3}$ is the constant isotropy generator as above and
$\La_{1}$ and $\La_{2}$ are functions of $r$ and $t$ that satisfy
\leqn{wang}{ 
[\La_{2},\La_{3}]=\La_{1} \AND
[\La_{3},\La_{1}]=\La_{2}\;. 
}

In this paper we will only consider the \emph{static magnetic case} for which
$\La_{1}$ and $\La_{2}$ as well as $N$ and $S$ are functions of $r$
only and the `electric' part $\tilde{A}$ of the potential vanishes.

With
\leqn{lapm}{ \Lpm := \mp \La_{1} - i \La_{2} } 
equations \eqref{wang} become 
\leqn{nwang}{ [\Lo,\Lpm] = \pm 2 \Lpm }
and the $\Lpm(r)$ satisfy the reality condition
\leqn{lam}{ \Lmp = -c(\Lpm) }

With this choice for the gauge potential, the Einstein-Yang-Mills (EYM)
equations can be written as \cite{k5281}
\lgath{feq}{
m' = NG + r^{-2}P,                         \label{feq1}\\
S^{-1}S' = 2 r^{-1}G,                        \label{feq2}\\
r^{2}N \Lp'' + 2(m-r^{-1}P)\Lp' + \Fc = 0,   \label{feq3}\\
[\Lp',\Lm] + [\Lm',\Lp] = 0                  \label{feq4}
}
where ${}' := d/dr$ and
\lgath{vardefs}{
N   =: 1 - \frac{2m}{r},  \quad
G   := \half (\Lp',\Lm'), \quad P := -\half (\Fh,\Fh), \label{vardefs1} \\
\Fh := \ihalf(\Lo-[\Lp,\Lm]),      \label{vardefs2} \\
\Fc := -i[\Fh,\Lp].                \label{vardefs3}
}
Here $(\:,\:)$ is an invariant inner product on $\g$. It is determined up to a
factor on each simple component of a semi-simple $\g$ and induces a norm $|.|$
on (the Euclidean) $\h$ and therefore its dual. We choose these factors so
that $(\:,\:)$ restricts to a negative definite inner product on $\g_0$.

For several purposes, in particular for numerical solutions, equations
\eqref{feq1}-\eqref{feq3} are best replaced by an equivalent system that
regularizes the almost singularity when $N$ is close to zero. These equations,
introduced by Breitenlohner, Forg\'acs and Maison \cite{k5428} in the $SU(2)$
case, take the form
\lgath{feqBFM}{
\dot{r}   = r\Nc,\quad \dot{\Nc} = \Nc(\Kc-\Nc) - 2\Gc,\quad
\dot{\Kc} = 1-\Kc^{2}+2\Gc,\quad \dot{\Sc} = (\Kc-\Nc)\Sc\label{feqBFM1}\\
\dot{\Lp} = r U_{+} \AND \dot{U}_{+} = -(\Kc-\Nc)U_{+}-\Fc/r\label{feqBFM2} 
}

where
\leqn{bfmDef}{
\Nc:=\sqrt{N},\;\Sc:=\Nc S,\;\Kc:=\frac{1}{2\Nc}\left(1+N+2G-2P/r^{2}\right),\; U_{+}:=\Nc\Lp',\;\Gc:=\half\|U_{+}\|^{2}
}

and the dot denotes a $\tau$-derivative. Note that the $\tau$ variable
behaves somewhat like the logarithm of $r$ near $0$ and for
$r\ra\infty$ (where $\Nc\ra1$).

For later use, we introduce a non-degenerate Hermitian inner product
$\g$ by
$\hip{X}{Y} := -(c(X),Y)$ for all $X$,$Y$ in $\g$. 
% recalling that $c:\g \rightarrow \g$ is the conjugation operator
% determined by the compact real form $\g_{0}$.
Then $\hip{\;\,}{\;}$ restricts to a real positive definite
inner product on $\g_{0}$. From the invariance properties of $(\;,\;)$ it
follows that  $\hip{\;\,}{\;}$ satisfies
\eqn{hiprel}{
\hip{X}{Y}  = \overline{\hip{Y}{X}} \; , \quad
\hip{c(X)}{c(Y)}  =  \overline{\hip{X}{Y}} \; , \AND
\hip{[X,c(Y)]}{Z} = \hip{X}{[Y,Z]} \;
}
for all $X,Y,Z \in \g$. Treating $\g$ as a $\mathbb{R}$-linear space
by restricting scalar multiplication to multiplication by reals,
we can introduce a positive definite inner product
$\rip{\;\,}{\;}: \g\, \times \,\g \rightarrow \mathbb{R}$
on $\g$ defined by
\leqn{ripdef}{
\rip{X}{Y} := \text{Re}\hip{X}{Y} \quad \forall \; X , Y \in \g \; .
}

Let $\norm{\;}$ denote the norm induced on $\g$ by $\rip{\;}{\;}$, i.e.
\leqn{norm}{
\norm{X} := \sqrt{\rip{X}{X}} \quad \forall \; X\in \g \; .
}

From the above properties satisfied by $\hip{\;\,}{\;}$, it
straighforward to verify that $\rip{\;\,}{\;}$ satisfies
\leqn{rip}{
\rip{X}{Y} = \rip{Y}{X}, \quad
\rip{c(X)}{c(Y)}  =  \rip{X}{Y} , \AND
\rip{[X,c(Y)]}{Z}   = \rip{X}{[Y,Z]} 
}
for all $X,Y,Z \in \g$.

Using the norm \eqref{norm}, equations \eqref{vardefs1}
can be written as
\leqn{vardefsa}{
G   = \half \|\Lp'\|^{2} \AND P = \half \|\Fh\|^{2}  
}

which shows that $G\ge0$ and $P\ge0$. 
Energy density, radial and tangential pressure are then given by
\leqn{ener}{
 4\pi e          = r^{-2}(NG+r^{-2}P), \quad
 4\pi p_{r}      = r^{-2}(NG-r^{-2}P), \quad
 4\pi p_{\theta} = r^{-4}P.            
}

The main result of this paper is that the EYM equations
\eqref{feq1}-\eqref{feq4} admit local bounded solutions in the neighborhood of
the origin $r=0$, a black hole horizon $r=r_{H}>0$, and as $r\ra\infty$.  To
prove this local existence to the EYM equations \eqref{feq1}-\eqref{feq4}, we
proceed in three steps.
\begin{enumerate}
\item First we prove the existence of local solutions 
$\{\Lp(r),m(r)\}$ to the equations 
\eqref{feq1} and  \eqref{feq3}.
\item Then we determine which solutions from step 1 
satisfy equation $\eqref{feq4}$.
\item Finally,  equation \eqref{feq2} can be integrated for
all  solutions from step 2 to obtain the metric function $S(r)$.
\end{enumerate}

Our method for carrying out the first step will be to prove that there
exists a change of variables so that the field equations \eqref{feq1} and
\eqref{feq3} can be put into a form to which the following (slight
generalization of a) theorem by Breitenlohner, Forg\'acs and Maison
\cite{k5428} applies.

\begin{thm} \mlabel{BFM}

The system of differential equations
\lalign{bfmeq}{
t \frac{du_{i}}{dt} &= t^{\mu_{i}}f_{i}(t,u,v) & i=1,\ldots,m \\
t \frac{dv_{j}}{dt} &= -h_{j}(u)v_{j} + t^{\nu_{j}}g_{j}(t,u,v) & j=1,\ldots,n
}

where $\mu_{i}$, $\nu_{j}$ are integers greater than 1, $f_{i}$ and $g_{j}$
analytic functions in a neighborhood of $(0,\mathbf{c}_{0},0)\in \RE^{1+m+n}$,
and $h_{j}:\RE^{m}\ra\RE$ functions, positive in a neighborhood of
$\mathbf{c}_{0}\in\RE^{m}$, has a unique analytic solution
$t\mapsto(u_{i}(t),v_{j}(t))$ such that 
\leqn{bfmsol}{ u_{i}(t) = c_{i} + O(t^{\mu_{i}}) \AND v_{j}(t) =
  O(t^{\nu_{j}}) }

for $|t|<R$ for some $R>0$ if $|\mathbf{c}-\mathbf{c}_{0}|$ is small enough.
Moreover, the solution depends analytically on the parameters $c_{i}$.
\end{thm}

The next lemma shows that if $\{\Lp(r),m(r)\}$ is a solution
to the field equations \eqref{feq1} and  \eqref{feq3} then the quantity 
$[\Lp',\Lm] + [\Lm',\Lp]$ satisfies a first order linear
differential equation. This unexpected result is what
allows us to carry out step 2 and thereby construct local
solutions.

\begin{lem}\label{constraint}\mnote{constraint}
If $\{\Lp(r),m(r)\}$ is a solution
to the field equations \eqref{feq1} and  \eqref{feq3} then
$\gamma(r) := [\Lp(r),\Lm'(r)] + [\Lm(r),\Lp'(r)]$
satisfies the differential equation
$\gamma' = -2(r^{2}N)^{-1}\left(m-r^{-1}P\right)\gamma$ .
\end{lem}
\begin{proof}
Differentiating $\gamma$ yields 
\alin{constraint3}{
    \gamma'&= [\Lp,\Lm'']+[\Lm,\Lp''] && \\
    & = -\frac{2}{r^{2}N}\left(m-\frac{1}{r}P\right)\gamma
    +\frac{i}{r^{2}N}([\Lm,[\Fh,\Lp]]+[\Lp,[\Lm,\Fh]])\;, && \text{by
      \eqref{feq1} and \eqref{feq3}} 
} 
while $[\Lm,[\Fh,\Lp]]+[\Lp,[\Lm,\Fh]] = 0$
by \eqref{nwang}, \eqref{vardefs2}, and the Jacobi identity. 
Combining the these two results proves the lemma.
\end{proof} 

\sect{cmpt}{Solving the field equations computationally}

Here we describe without proofs the more practical aspects of solving the
field equations. That all these constructions work will follow from the proofs
given in sections \ref{alg} and \ref{lep}.

We consider only situations where the EYM field is nonsingular at the
center and/or the gravitational and \YM\ field fall off rapidly at
infinity. From the expressions for the physical quantities
\eqref{ener} and \eqref{vardefsa} it then follows that $\Fh$ vanishes
there so that by \eqref{vardefs2} $\{\Lo,\Lp,\Lm\}$ form a triple
defining an $\sL(2)$ (or $A_{1}$) subalgebra of $\g$. This restricts
the (constant) $\Lo$ vector to be a \emph{defining vector} of such a
subalgebra which have been classified by Dynkin
\cite{k4779}. Alternatively, in the terminology of \cite{k6494}, the
set $\{\Lo,\Op,\Om\}$, where we define $\Opm$ to be the limiting
values of $\Lpm$ at $r=0$ or $\infty$, is called a standard triple
defining a nilpotent orbit, $\Lo$ is the neutral and $\Op$ the
nilpositive element.

It is known \cite{k4779} that there is always an automorphism of $\g$ that
maps $\Lo$ onto the fundamental Weyl chamber so that its characteristic (or
weighted Dynkin diagram)
$\chi=(\chi_{1},\ldots,\chi_{\ell}):=(\alpha_{1}(\Lo),\ldots,\alpha_{\ell}(\Lo))$
with respect to a chosen Cartan subalgebra and its dual basis
$\{\alpha_{1},\ldots,\alpha_{\ell}\}$, satisfies $\chi_{k}\in\{0,1,2\}
\;\forall\;k=1,\ldots,\ell$. We call the symmetry group action and the vector
$\Lo$ \emph{regular} if $\Lo$ lies in the interior of the Weyl chamber or,
equivalently, if all $\chi_{k}$ are positive. The action and $\Lo$ are called
\emph{principal} if $\chi_{k}=2\:\forall k$.

Since for semi-simple gauge groups all constructions are easily decomposed
into those of the simple factors a classification need only be done for the
simple Lie groups. It turns out that only for the $A_{\ell}$ (or
$\sL(\ell+1)$) series of the classical Lie algebras and then only for even
$\ell$ there are regular actions other than the principal one. (See
\cite{eymg}, Theorem 2.)  Moreover, in those cases the gauge field corresponds
to one of a direct sum of two lower-dimensional simple Lie algebras of
type $A_{\ell}$. It is for that reason that we could confine ourselves in
\cite{eymg} to the principal case when studying the local existence problem of
solutions for regular symmetry group actions.

In general, given a fixed defining vector $\Lo$, the \YM\ field will be fully
determined, in view of \eqref{lam}, if $\Lp$ is known as a function of $r$.
Condition \eqref{nwang} implies that $\Lp$ must lie in the vector subspace
$V_{2}$ of $\g$ where we define, more generally, the eigenspaces of $\ad(\Lo)$
by
\eqn{Vn}{
V_{n} := \{ X\in \g \; | \; \Lo . X = n X \}  \quad n\in \mathbb{Z} \; ,
}

Here, and in the following, the adjoint action of $\cspan{\{\Lo,\Op,\Om\}}
\cong \Sl{2}$ on $\g$, is denoted by a dot,
\eqn{ad}{
X.Y := \ad(X)(Y) \qquad \forall\; \text{$X\in \cspan{\{\Lo,\Op,\Om\}}$, $ Y\in\g$.}
}

In terms of a Chevalley-Weyl basis (see \cite{k5157} or \cite{eymg} for the
notation) $\{ \hb_{j}:=\hb_{\alpha_{j}}, \eb_{\alpha}, \eb_{-\alpha} |
j=1,\ldots,\ell; \alpha\in R^{+} \}$, where $R^{+}$ is the set of positive
roots of $\g$ with respect to the root basis
$\{\alpha_{1},\ldots,\alpha_{\ell}\}$, we then have
\leqn{lap}{
\Lp(r) = \sum_{\alpha\in \So} w_{\alpha}(r) \eb_{\alpha} \in V_{2} \AND 
\Lm(r) = \sum_{\alpha\in \So} \bar{w}_{\alpha}(r)  \eb_{-\alpha} \in V_{-2}
}

where
\leqn{Slam}{ \So := \{\alpha\in R\, | \, \alpha(\Lo)=2 \}. }

In the regular case the Stiefel set $\So$ is necessarily a $\Pi$-system
\cite{k4779}, i.e. forms the basis of a root system generating a
subalgebra $\g_{\Lo}$ of $\g$ (see \cite{k4295}). Since for base root
vectors $\alpha, \beta$ also $[\eb_{\alpha},\eb_{\beta}] \neq 0$ only
if $\beta=-\alpha$ and $[\eb_{\alpha},\eb_{-\alpha}]=\hb_{\alpha}$ it
follows then from \eqref{vardefs2} that $\Fh$ lies in the Cartan
subalgebra of $\g_{\Lo}$.  Substituting \eqref{lap} into \eqref{feq4}
leads to \leqn{constr}{
w'_{\alpha}\bar{w}_{\beta}-w_{\alpha}\bar{w}'_{\beta}=0 \;\forall
\alpha,\beta\in \So }

so that all the phases of the complex $w_{\alpha}(r)$ are constant.

In the general case, $\So$ will not be linearly independent and the set
$\{[\eb_{\alpha},\eb_{-\beta}]\;|\;\alpha,\beta\in\So\}$ will not lie in the
Cartan subalgebra $\h$ of $\g$. There is no simple gauge transformation now
that will allow us to choose the $w_{\alpha}$ real, although, as follows from
Lemma \ref{constraint}, if \eqref{feq4} is satisfied at one regular value of
$r$ it will be in a whole interval.

The \YM\ field $\Fh$ no longer takes all its values in the Cartan subalgebra.
It is instead given by
\leqn{YMf}{
\Fh = \frac{i}{2}\left( \Lo -{\DS\sum_{\alpha,\beta\in\So}} w_{\alpha}(r)\bar{w}_{\beta}(r)[\eb_{\alpha},\eb_{-\beta}]\right).
}

However, the term $\Fc$ appearing in the \YM\ equation \eqref{feq3}
lies in $V_{2}$ since $[X,[Y,Z]]\in V_{2}$ whenever two of the
$X$,$Y$,$Z$ lie in $V_{2}$ and one in $V_{-2}$.

For computational purposes \eqref{feq3} can then be written
\leqn{feq3n}{
r^{2}N w''_{\alpha} + 2(m-r^{-1}P)w'_{\alpha}+ f_{\alpha} = 0
}

where \leqn{deff}{ 
f_{\alpha} := w_{\alpha} +
\frac{1}{2}{\DS\sum_{\beta,\gamma,\delta\in\So}}
\mu_{\alpha\beta\gamma\delta}w_{\beta}w_{\gamma}\bar{w}_{\delta}
}

and \leqn{defP}{ 
P = \frac{1}{8}\Vert\Lo\Vert^{2} -
{\DS\sum_{\alpha\in\So}}|\alpha|^{-2}|w_{\alpha}|^{2} -
\frac{1}{4}{\DS\sum_{\alpha,\beta,\gamma,\delta\in\So}}
|\alpha|^{-2}\mu_{\alpha\beta\gamma\delta}\bar{w}_{\alpha}w_{\beta}w_{\gamma}\bar{w}_{\delta}.
}

(Here the Jacobi identity and the invariance of the inner product $(\:,\:)$
were used.) Thus the general structure of the Lie algebra $\g$ enters the
equations only via the quantity $\mu_{\alpha\beta\gamma\delta}$ which is
defined by 
\leqn{defmu}{ 
[\eb_{\beta},[\eb_{\gamma},\eb_{-\delta}]] =
  \sum_{\alpha\in\So}\mu_{\alpha\beta\gamma\delta}\,\eb_{\alpha}
  \;\;\forall\;\beta,\gamma,\delta\in\So.  
}

To start off the numerical integration of a bounded solution near one of the
singular points we need to sum a power series at a point nearby. This is again
similar to the method used in the regular case, but we must allow for possibly
complex functions $w_{\alpha}(r)$. This has most of all the unpleasant effect
that the values that the $w_{\alpha}$ can take at the center and at infinity
form no longer just a finite set with a few signs to be chosen arbitrarily, but
an $M$-dimensional real variety in the space $\CO^{M}$ (where $M$
denotes the number of elements in $\So$ and thus the number of functions
$w_{\alpha}$ needed to characterize the gauge potential $\Lp$).  Since there
is no reason to believe that for a global solution on $[0,\infty)$ the values
of $\Lp(0)$ and $\Lp(\infty)$ should be the same we can, for example, choose
an arbitrary $\Op=\Lp(\infty)$ which amounts to a global gauge choice. But
then the value of $\Lp$ at the center could be any point of this
$M$-dimensional real variety so that the coordinates describing the latter
need to be given as parameters as well as some of the first few power series
coefficients.

Wishing to find a local analytic solution to equations \eqref{feq1} and
\eqref{feq3n} we expand all quantities in a power series in $r$ near $r=0$.
The lowest order terms $w_{\alpha,0}=w_{\alpha}(0)$ are constrained by $\Fh=0$
so that
\leqn{wconstr}{
\Lo =  [\Op,\Om] \AND  \Om=-c(\Op)\quad\text{or}
}

or
\leqn{wconstr1}{
\Lo =: \sum_{i=1}^{\ell}\lambda_{i}\hb_{i}=\sum_{\alpha,\beta\in\So} w_{\alpha,0}\bar{w}_{\beta,0}[\eb_{\alpha},\eb_{-\beta}]
} 

which gives, if we write $\hb_{\alpha}=\sum_{i=1}^{\ell}h_{\alpha,i}\hb_{i}$
and $R$ denotes the set of all roots of the Cartan subalgebra $\h$ of $\g$,
\leqn{wconstr2}{
\sum_{\alpha\in \So}h_{\alpha,i} |w_{\alpha,0}|^{2} = \lambda_{i}\;\; 
(i=1,\ldots,\ell) \AND \sum_{\substack{\alpha,\beta\in\So\\
\alpha-\beta\in R}}w_{\alpha,0}\bar{w}_{\beta,0}[\eb_{\alpha},\eb_{-\beta}] 
= 0.
}

We have not yet been able to determine whether equations \eqref{wconstr2} have
solutions for all simple Lie algebras and all defining vectors $\Lo$. For all
low-dimensional examples we have so far considered solutions exist and form an
$M$-dimensional real variety in $\CO^{M}$. Moreover, it appears that there
always exist vectors $\Op$ with only real components with respect to the basis
$\{\eb_{\alpha}\}$.

Suppose now that $\Op$ has been chosen. Then equation \eqref{feq1} yields the
recurrence relation
\leqn{powserM}{
m_{k} = \frac{1}{k}\left(P_{k+1}+G_{k-1}-2{\DS\sum_{i=2}^{k-3}}m_{k-i}G_{i}\right)
}

and \eqref{feq3} gives
\leqn{powserL}{
 \Ab(\Lambda_{+,k}) - k(k-1)\Lambda_{+,k} = \mathbf{b}_{k} \quad (k=2,3,\ldots)
}

where the $\mathbf{b}_{k}$ are complicated expressions of lower order terms 
and $\Ab$ is defined by
\leqn{Adef}{
\Ab = \half \ad(\Op) \circ \bigl(\,\ad(\Om) + \ad(\Op)\circ c\,\bigr).
}

This $\Ab$ is an $\RE$-linear operator which turns out to be symmetric with
respect to the inner product \eqref{ripdef} and restricts to $V_{2}$.
As will be shown in section \ref{alg}, half of its eigenvalues are
zero and the remaining ones are positive integers of the form $k(k+1)$ for
$k=1,2,\ldots$. We will later need the notation
\leqn{Edef}{
  \Ec := \{ s\in\NA \:|\: s(s+1) \text{\ is an eigenvalue of\ } \Ab \}
}

Moreover, to every eigenvector with nonzero eigenvalue there exists one 
(its multiple by $i$) that has eigenvalue zero.

Explicitly, if we write $\Op = \sum_{\alpha=1}^{M}(u_{\alpha,0}+i v_{\alpha,0})\eb_{\alpha}$, where $M:=|\So|$, then $\Ab$ is given by the matrix
\leqn{Amat}{
(\Ab) = \begin{pmatrix}a & b\\c & d\end{pmatrix}
}

where
\lalign{abcd}{
 a_{\alpha\beta} &= -\sum_{\gamma,\delta}\left(\mu_{\alpha\delta(\beta\gamma)}u_{\gamma,0}u_{\delta,0} +\mu_{\alpha\delta[\beta\gamma]}v_{\gamma,0}v_{\delta,0}\right) \\
 b_{\alpha\beta} &= \phantom{-}\sum_{\gamma,\delta}(\mu_{\alpha\delta[\beta\gamma]}-\mu_{\alpha\gamma(\beta\delta)})u_{\gamma,0}v_{\delta,0} \\
 c_{\alpha\beta} &= \phantom{-}\sum_{\gamma,\delta}(\mu_{\alpha\gamma[\beta\delta]}-\mu_{\alpha\delta(\beta\gamma)})u_{\gamma,0}v_{\delta,0} \\
 d_{\alpha\beta} &= -\sum_{\gamma,\delta}\left(\mu_{\alpha\gamma[\beta\delta]}u_{\gamma,0}u_{\delta,0} +\mu_{\alpha\gamma(\beta\delta)}v_{\gamma,0}v_{\delta,0}\right)
}

Now let $\Eb$ be the matrix whose columns are eigenvectors of $\Ab$ such that
\leqn{AmatEV}{
 \Ab \Eb = \Eb \La
}

where $\La=\diag(0\ldots0 \lambda_{1}\ldots\lambda_{M})$ with
$\lambda_{i}\leq\lambda_{j}$ if $i\leq j$. Then we know (from section
\ref{alg}) that if $\Eb$ is of the form $\Eb=\begin{pmatrix}\ast & e_{1}\\ 
  \ast & e_{2}\end{pmatrix}$ then also $\begin{pmatrix}-e_{2} & e_{1} \\ e_{1}
  & e_{2}\end{pmatrix}$ will be a matrix of eigencolumnvectors with now the
correct correspondence between the real and imaginary parts.

To generate the powerseries for $u$ and $v$ we now find from the recurrence
relation \eqref{powserL} for the coefficients of $r^{k}$ that
\leqn{powserS}{
 \begin{pmatrix} u_{k} \\ v_{k} \end{pmatrix} = \Eb^{T}X_{k}
}

where
$X_{k,\alpha}=\bigl((\Eb^{T})^{-1}\mathbf{b}_{k}\bigr)_{\alpha}/(\lambda_{\alpha}-k(k-1))$
if $\lambda_{\alpha}\neq k(k-1)$ or a new free parameter otherwise.

This shows that the general solution near $r=0$ is of a similar form as in the regular case (\cite{eymg}, Theorem 4), namely
\leqn{powserUV}{ w_{\alpha}(r) = w_{\alpha,0} +
  \sum_{\beta=1}^{M}\Eb_{\alpha\beta}r^{k_{\beta}+1}\tilde{w}_{\beta}(r),\quad
  \alpha=1,\ldots,M }

where $k_{\alpha}(k_{\alpha}+1)$ is the $\alpha$-th nonzero eigenvalue
of $\Ab$ and $\tilde{w}_{\alpha}(r)$ are analytic functions near
$r=0$. The general solution $\{m(r),w_{\alpha}(r)|\alpha=1,\ldots,M\}$
is determined by $M$ real parameters for the initial values
$w_{\alpha}(0)$ as well as another $M$ real parameters that can be
arbitrarily chosen in the coefficients $\tilde{w}_{\beta}(0)$ of
$r^{k_{\beta}+1}$.

The general power series solution in $r^{-1}$ near infinity is very similar.

For very special choices of the parameters one expects to find
solutions of EYM-equations correesponding to gauge groups that are
subgroups of $G$. In one simple case this is easy to see.

Since every compact semi-simple (non-Abelian) Lie group
\mnote{semi-simple or non-Abelian?} has $SU(2)$ as a subgroup there
must be $SU(2)$-solutions embedded among general $G$-solutions for any
symmetry group action. There may be several conjugacy classes of such
subgroups, but there is a distinguished one related to the
homomorphism defined by the action. Let the action and hence $\Lo$ be
given and pick any $\Op$ thus selecting a specific $\sL(2)$-subalgebra
generated by the triple $\{\Lo,\Op,\Om=-c(\Op)\}$ among the conjugacy
class associated with $\Lo$. If we then let \leqn{specS}{ \Lp(r) =
\mathtt{v}(r)\Op \quad \text{with\ }\mathtt{v}(0)=1 }

it follows from \eqref{feq4} that $\mathtt{v}(r)=e^{i\gamma_{0}}u(r)$ for a
constant $\gamma_{0}$ and a real function $u(r)$. The remaining equations
\eqref{feq1}-\eqref{feq3} then become
\lgath{specE}{
m' = g_{0}^{2}\bigl(N{u'}^{2} + \half r^{-2}(1-u^{2})^{2}\bigr), \label{specE1}\\
S^{-1}S' = 2 g_{0}^{2} r^{-1}{u'}^{2},         \label{specE2}\\
r^{2}N u'' + \bigl(2m-g_{0}^{2}r^{-1}(1-u^{2})^{2}\bigr)u' + (1-u^{2})u = 0,   \label{specE3}
}

where $g_{0}:=\half\Vert\Lo\Vert$. They reduce with
\leqn{nvar}{
\rho := r/g_{0}, \quad \mu := m/g_0
}

to the equations for the $SU(2)$ theory,
\lgath{specEE}{
d\mu/d\rho = N (du/d\rho)^{2}+\half \rho^{-2}(1-u^{2})^{2}, \label{specEE1}\\
S^{-1}dS/d\rho = 2 r^{-1}(du/d\rho)^{2},         \label{specEE2}\\
\rho^{2}Nd^{2}u/d\rho^{2} + \bigl(2\mu-\rho^{-1}(1-u^{2})^{2}\bigr)du/d\rho 
+ (1-u^{2})u = 0,   \label{specEE3}
}

where now $N=1-2\mu/\rho$.

Much is known about the solutions of these equations. In particular, it
follows, that for any compact semi-simple Lie group $G$ and any action of the
symmetry group $SU(2)$ an infinite discrete set of global asymptotically flat
(soliton and black hole) solutions exists.

\sect{class}{The spherically symmetric static EYM models for some small gauge
groups}

We have not yet found a simple general way to derive the field
equations for arbitrary semi-simple $\g$ and arbitrary choice of
$\Lo$. But in the following tables \ref{ABtab} to \ref{FGtab} we list
for lower dimensional Lie groups and the actions of $SU(2)$ given by
their characteristic some basic properties of the system of equations,
namely the size of the size $M=|S_{\Lo}|$ of the Stiefel set which
corresponds to the number of complex functions $w_{\alpha}(r))$ that
describe the gauge potential, the set $\Ec$ with the superscript
denoting the dimension of the eigenspace if it is greater than one,
and the subalgebra to which the equations reduce if $S_{\Lo}$ is a
$\Pi$-system (`-' indicates that there is no reduction). We leave out
the principal action, which is always regular, as well as the trivial
action of the symmetry group.

Just to give an idea of the structure the $M$-dimensional real
subvariety $\Sigma$ of $\CO^M$ we list in table \ref{initV} the
equations that define it for a few cases. (In the regular case for a
Lie algebra of rank $\ell$ the equations would require that
$|w_i|=\const$ for certain fixed constants for all
$i=1,2,\ldots,\ell$.)

When one wishes to find global numerical solutions one can, for
example, pick a simple choice for $\Lp(\infty)$ -- it seems possible
to choose all the $w_{\alpha}(\infty)$ real, and many of them 0. But
then the data at $r=0$ must be left arbitrary, i.e. a suitable
parametrization of the variety $\Sigma$ must be introduced. This is
not difficult for the low dimensional cases, but not straightforward.

\begin{center}
%{\small
\begin{table}[htbp]\label{ABtab}
\caption{Irregular actions for Lie algebras 
$A_2$ to $A_4$ and $B_2$ to $B_4$}
$$
\begin{array}{||c|c|c|c||c|c|c|c||}
\hline\hline
%\rule[-0.5ex]{0pt}{2.6ex}
\chi & |S_{\Lo}| & \Ec & \text{reduction} &
\chi & |S_{\Lo}| & \Ec & \text{reduction} \\
\hline\hline
\multicolumn{4}{||c||}{A_{2}\quad (SU(3))}& 
\multicolumn{4}{c||}{B_{2}\quad (SO(5))}\\
\hline\hline
1 1 & 1 & 1 & A_{1} &
0 1 & 1 & 1 & A_{1} \\
    &   &   &       &
2 0 & 3 & 1^3 & -   \\
\hline\hline
\multicolumn{4}{||c||}{A_{3}\quad (SU(4))}&
\multicolumn{4}{c||}{B_{3}\quad (SO(7))}\\
\hline\hline
1 0 1 & 1 & 1   & A_{1} &
0 1 0 & 1 & 1   & A_{1} \\
\hline
0 2 0 & 4 & 1^{4} & - &
1 0 1 & 2 & 1^{2} & A_{1}\oplus A_{1} \\
\hline
2 0 2 & 4 & 1^{3}, 6  & - &
2 2 0 & 4 & 1,2^{2},3 & - \\
\hline
    &   &   &       &
2 0 0 & 5 & 1^{5} & - \\
\hline
    &   &   &       &
0 2 0 & 6 & 1^{5},2 & - \\
\hline\hline
\multicolumn{4}{||c||}{A_{4}\quad (SU(5))}&
\multicolumn{4}{c||}{B_{4}\quad (SO(9))}\\
\hline\hline
1 0 0 1 & 1 & 1          & A_{2} &
0 1 0 0 & 1 & 1 & A_{1} \\
\hline
2 1 1 2 & 3 & 1, 2, 3    & A_{3} &
2 1 0 1 & 3 & 1^{2}, 3 & B_{2}\oplus A_{1} \\
\hline
1 1 1 1 & 3 & 1^{2}, 2 & A_{3}\oplus A_{2} &
1 0 1 0 & 4 & 1^{4} & - \\
\hline
0 1 1 0 & 4 & 2^{4}     & - &
0 2 0 1 & 5 & 1,2^{3},3 & - \\
\hline
2 0 0 2 & 6 & 2^{5}, 3 & - &
2 2 2 0 & 5 & 1, 3^{3}, 5  & - \\
\hline
    &   &   &       &
0 0 0 1 & 6 & 1^{6} & - \\
\hline
    &   &   &       &
2 2 0 0 & 6 & 1,2^{4},3 & - \\
\hline
    &   &   &       &
2 0 0 0 & 7 & 1^{7} & - \\
\hline
    &   &   &       &
2 0 2 0 & 8 & 1^{4}, 2^{2}, 3^{2} & - \\
\hline
    &   &   &       &
0 0 2 0 & 9 & 1^{6},2^{3} & - \\
\hline
    &   &   &       &
0 2 0 0 & 10 & 1^{9},2 & - \\
\hline\hline
\end{array}
$$
\end{table}
%} \small
\end{center}

\begin{center}
\begin{table}[htbp]\label{CDtab}
\caption{Irregular actions for the Lie algebras $C_{3}$ ($Sp(6)$), $C_4$ ($Sp(8)$) and $D_4$ ($SO(8)$)}
$$
\begin{array}{||c|c|c|c||c|c|c|c||}
\hline\hline
\chi & |S_{\Lo}| & \Ec & \text{reduction} &
\chi & |S_{\Lo}| & \Ec & \text{reduction} \\
\hline\hline
\multicolumn{4}{||c||}{C_{3}\quad (Sp(6))}& 
\multicolumn{4}{c||}{}\\
\hline\hline
1 0 0 & 1 & 1 &  A_{1} &
    &   &   &       \\
\hline
2 1 0 & 2 & 1,3  & C_{2} &
    &   &   &       \\
\hline
0 1 0 & 3 & 1^{3} & - &
    &   &   &       \\
\hline
0 2 0 & 4 & 1,2^{3} & - &
    &   &   &       \\
\hline
2 0 2 & 5 & 1^{3},2,3 & - &
    &   &   &       \\
\hline
0 0 2 & 6 & 1^{6}& - &
    &   &   &       \\
\hline\hline
\multicolumn{4}{||c||}{C_{4}\quad (Sp(8))}& 
\multicolumn{4}{c||}{D_{4}\quad (SO(8))}\\
\hline\hline
1 0 0 0 & 1 & 1 & A_{1} &
0 1 0 0 & 1 & 1 & A_{1} \\
\hline
2 1 0 0 & 2 & 1,3 & C_{2} &
1 0 1 1 & 3 & 1^{3} & A_{1}\oplus A_{1}\oplus A_{1} \\
\hline
0 1 0 0 & 3 & 1^{3} & - &
0 2 0 2 & 5 & 1,2^{3},3 & - \\
\hline
2 2 1 0 & 3 & 1,3,5 & C_{3} &
0 2 2 0 & 5 & 1,2^{3},3 & - \\
\hline
2 0 1 0 & 5 & 1^{3},2,3    & - &
2 2 0 0 & 5 & 1,2^{3},3 & - \\
\hline
0 1 1 0 & 5 & 1^{2},2^{3} & - &
0 0 0 2 & 6 & 1^{6} & - \\
\hline
0 0 1 0 & 6 & 1^{6}& - &
0 0 2 0 & 6 & 1^{6} & - \\
\hline
2 2 0 2 & 6 & 1^{2},2,3^{2},5 & - &
2 0 0 0 & 6 & 1^{6} & - \\
\hline
0 2 0 2 & 7 & 1^{3},2,3^{3}  & - &
2 0 2 2 & 6 & 1^{3},2,3^{2} & - \\
\hline
0 2 0 0 & 8 & 1^{5},2^{3} & - &
0 2 0 0 & 8 & 1^{7},2 & - \\
\hline
2 0 0 2 & 9 & 1^{6},2^{2},3 & - &
 & & & \\
\hline
0 0 0 2 & 10 & 1^{10} & - &
& & & \\
\hline\hline
\end{array}
$$
\end{table}
%}
\end{center}

\begin{center}
%{\small
\begin{table}\label{FGtab}
\caption{Irregular actions for the Lie algebras $F_4$ and $G_{2}$}
$$
\begin{array}{||c|c|c|c||c|c|c|c||}
\hline\hline
\chi & |S_{\Lo}| & \Ec & \text{reduction} &
\chi & |S_{\Lo}| & \Ec & \text{reduction} \\
\hline\hline
\multicolumn{4}{||c||}{F_{4}}& 
\multicolumn{4}{c||}{G_2}\\
\hline\hline
1 0 0 0 & 1 & 1 & A_{1} &
0 1     & 1 & 1 & A_{1} \\
\hline
1 0 1 2 & 3 & 1,3,5 & C_{3} &
1 0     & 1 & 1 & A_{1} \\
\hline
1 0 1 0 & 5 & 1^{3},2,3 & - &
0 2 & 4 & 1^{3},2 & - \\
\hline
0 1 0 1 & 5 & 1^{2},2^{3} & - &
  & & & \\
\hline
0 1 0 0 & 6 & 1^{6} & - &
  & & & \\
\hline
2 0 0 1 & 6 & 1,2^{4},3 & - &
  & & & \\
\hline
2 2 0 2 & 6 & 1,2,3,5^{2},7 & - &
  & & & \\
\hline
0 0 0 1 & 7 & 1^{7}& - &
  & & & \\
\hline
2 2 0 0 & 7 & 1,3^{5},5 & - &
  & & & \\
\hline
0 0 0 2 & 8 & 1,2^{7} & - &
  & & & \\
\hline
0 2 0 2 & 8 & 1^{3},2,3,4,5^{2} & - &
  & & & \\
\hline
0 0 1 0 & 9 & 1^{6},2^{3} & - &
  & & & \\
\hline
0 2 0 0 & 12 & 1^{6},2^{4},3^{2} & - &
  & & & \\
\hline
2 0 0 0 & 14 & 1^{13},2 & - &
  & & & \\
\hline\hline
\end{array}
$$
\end{table}
\end{center}

\begin{center}
%{\small
\begin{table}[htbp]\label{initV}
\caption{The ``initial value'' surface $\Sigma$ for a few irregular 
actions for some Lie algebras}
$$
\begin{array}{||c|c|m{90mm}||}
\hline\hline
\text{Lie algebra} & \chi & \multicolumn{1}{|c||}{\Sigma} \\
\hline\hline
A_3 & 0 2 0 & $|w_1|^2+|w_2|^2=1$, $|w_3|=|w_2|$, $|w_4|=|w_1|$, 
$w_1\bar{w}_3=-w_2\bar{w}_4$, $w_1\bar{w}_2=-w_3\bar{w}_4$ \\
\hline
B_4 & 1 0 1 0 & $|w_{1}|^{2}+|w_{2}|^{2}=1$, $|w_{3}|=|w_{1}|$, 
$|w_{4}|=1$, $w_{2}\bar{w}_{3}=w_{1}\bar{w}_{2}$ \\
\hline
E_6 & 1 2 0 0 0 1 & $|w_{1}|^{2}+|w_{3}|^{2}+|w_{5}|^{2}=3$, 
$|w_{4}|=|w_{3}|$, $|w_{5}|=|w_{2}|$, $|w_{6}|=2$, $|w_{7}|=|w_{1}|$, 
$w_{1}\bar{w}_{2}=w_{5}\bar{w}_{7}$, $w_{1}\bar{w}_{3}=-w_{4}\bar{w}_{7}$, 
$w_{1}\bar{w}_{4}=-w_{3}\bar{w}_{7}$, $w_{1}\bar{w}_{5}=w_{2}\bar{w}_{7}$, 
$w_{2}\bar{w}_{3}=-w_{4}\bar{w}_{5}$, $w_{2}\bar{w}_{4}=-w_{3}\bar{w}_{5}$ \\
\hline
F_4 & 1 0 1 0 & $|w_{1}|^{2}+|w_{3}|^{2}=3$, 
$|w_{3}|^{2}+|w_{4}|^{2}+|w_{5}|^{2}=4$, 
$-|w_{2}|^{2}+2|w_{3}|^{2}+|w_{5}|^{2}=3$, 
$w_{1}\bar{w}_{3}+w_{2}\bar{w}_{4}=w_{4}\bar{w}_{5}$ \\
\hline
G_2 & 0 2 & $|w_{1}|^{2}+2|w_{2}|^{2}+|w_{3}|^{2}=2$, 
$|w_{2}|^{2}+2|w_{3}|^{2}+|w_{4}|^{2}=2$, 
$w_{1}\bar{w}_{2}+2w_{2}\bar{w}_{3}+w_{3}\bar{w}_{4}=0$ \\
\hline\hline
\end{array}
$$
\end{table}
%}
\end{center}

\sect{alg}{Algebraic Results}

In this section we collect all of the algebraic results needed to
prove the local existence theorems. We will employ the same
notation as in \cite{eymg} section 6.

Before proceeding, we recall some results from \cite{eymg}.
\begin{prop} \label{hwv} \mnote{[hwv]}
There exists $M$ highest weight vectors
${\xi^{1},\xi^{2},\ldots,\xi^{M} }$ for the
adjoint representation of {\rm $\q$} on $\g$
that satisfy
\begin{itemize}
\item[(i)] the  $\xi^{j}$ have weights $2k_{j}$ where $j=1,2,\ldots,M$ and
$1 = k_{1} \leq k_{2} \leq \cdots \leq k_{M}$,
\item[(ii)] if $V(\xi^{j})$ denotes  the irreducible submodule of $\g$ 
generated by
$\xi^{j}$, then the  sum $\sum_{j=1}^{M} V(\xi^{j})$ is direct,
\item[(iii)] if $\xi^{j}_{l} = (1/l!)\Om^{l} . \xi^{j}$ then
\leqn{xiconj}{
c(\xi^{j}_{l}) = (-1)^{l} \xi^{j}_{2k_{j}-l} \; ,
}

\item[(iv)] $M = |S_{\lambda}|$ and the set 
$\{\xi^{j}_{k_{j}-1} \; | \; j= 1,2,\ldots M\}$ forms a basis for 
$V_{2}$ over $\mathbb{C}$.
\end{itemize}
\end{prop}

According to Lemma 1 of \cite{eymg} the $\mathbb{R}$-linear operator 
$\Ab : \g \rightarrow \g$ defined by \eqref{Adef} satisfies
\leqn{AV}{
\Ab(V_{2}) \subset V_{2} 
}

and hence restricts to an operator on $V_{2}$ which we denote by $A_{2}$.

We label the integers $k_{j}$ from proposition \ref{hwv} as follows
\begin{multline*}
1 = k_{J_{1}} = k_{J_{1}+1} = \cdots =  k_{J_{1}+m_{1}-1}
<  k_{J_{2}} = k_{J_{2}+1} = \cdots =  k_{J_{2}+m_{2}-1} \\
< \cdots <  k_{J_{I}} = k_{J_{I}+1} = \cdots =  k_{J_{I}+m_{I}-1} \; ,
\end{multline*}
where
$J_{1} = 1$, $\;J_{l} + m_{l} = J_{l+1}$ for $l = 1,2,\ldots, I$
and $J_{I+1} = M-1$. Then
\eqn{kf}{
\Ec = \{ \ke_{l} := k_{J_{l}} | l=1,2,\ldots, I\}\; .
}

The set $\{\xi^{j}_{k_{j}-1} \; | \; j= 1,2,\ldots M\}$ forms a basis over
$\mathbb{C}$ of $V_{2}$ by proposition \ref{hwv} (iv) while $I$ is the number
of distinct nonzero eigenvalues of $A_{2}$.\mnote{Is this correct?} Therefore
the set of vectors $\{X^{l}_{s},Y^{l}_{s} \;|\; l=1,2,\ldots,I \; ; \;
s=0,1,\ldots, m_{l}-1 \}$ where
\leqn{Xls}{
X^{l}_{s} := \left\{
\begin{array}{l}
\xi^{J_{l}+s}_{\ke_{l}-1} \quad \text{ if $\ke_{l}$ is odd} \\
i\xi^{J_{l}+s}_{\ke_{l}-1} \quad  \text{ if $\ke_{l}$ is even}
\end{array}
\right. \AND
Y^{l}_{s}  := i X^{l}_{s} \; ,
}

forms a basis of $V_{2}$ over $\mathbb{R}$. Then lemma 2 of
\cite{eymg} shows that this basis is an eigenbasis of
$V_{2}$ and we have
\leqn{A2x}{
A_{2}(X^{l}_{s})  = \ke_{l}(\ke_{l}+1) X^{l}_{s} \AND
A_{2}(Y^{l}_{s})  = 0 \text{\ for\ } l=1,2,\ldots,I \;\; 
s=0,1,\ldots,m_{l}-1 \; .
}

An immediate consequence of this result is that $\text{spec}(A_{2}) = \{0\}
\cup \{ \ke_{j}(\ke_{j}+1) \; | \; j = 1,2,\ldots I \}$ and $m_{j}$ is the
dimension of the eigenspace corresponding to the eigenvalue
$\ke_{j}(\ke_{j}+1)$. Note that $I$ is the number of distinct positive
eigenvalues of $A_{2}$.

Define 
\leqn{El}{
E^{l}_{0}  = \rspan{\{Y^{l}_{s}\;|\; s =0,1,\ldots, m_{l}-1\}} \; , \quad
E^{l}_{+}  = \rspan{\{X^{l}_{s}\;|\; s =0,1,\ldots, m_{l}-1\}} \; ,
}

and
\leqn{E0}{
E_{0}  = \bigoplus_{l=1}^{I} E^{l}_{0} \; , \qquad
E_{+}  = \bigoplus_{l=1}^{I} E^{l}_{+} \; .
}

Then
\leqn{kerA2}{ 
E_{0} = \ker\left(A_{2}\right)
}

and $E^{l}_{+}$ is the eigenspace of $A_{2}$ corresponding to the eigenvalue
$\ke_{l}(\ke_{l}+1)$.  Moreover, using proposition \ref{hwv} (iv), it is clear
that 
\leqn{V2}{ V_{2} = E_{0}\oplus E_{+} \; .  } 

To simplify notation in what
follows, we introduce one more quantity \leqn{Esl}{ E^{l} :=
  \bigoplus_{q=0}^{l} E^{q}_{0}\oplus E^{q}_{+}\: .  } We then have the useful
lemma from \cite{eymg}.
\begin{lem} \label{proja} \mnote{[proja]}
If $X\in V_{2}$ then
$X\in E^{l}$ if and only if
$\Op^{\ke_{l}} .\, X = 0$ or
$\Op^{\ke_{l}+2} .\, c(X) = 0$.
\end{lem}

We will also need the the map
$\;\; \widetilde{\;} : \mathbb{Z}_{\geq -1}  \rightarrow \{1,2,\ldots,I\}$
defined by
\leqn{tildemap}{
\text{ $\widetilde{-1} = \tilde{0} = 1$ and  
$\tilde{s} = \max\{ \;l\;|\;\ke_{l} \leq s \; \}$ if $s > 0$. }
}

It is shown in lemma 5 of \cite{eymg} that this map satisfies
\leqn{tildeprop}{
\text{$\ke_{\tilde{s}} \leq s$ for every
$s \in \mathbb{Z}_{\geq 0}$ and 
$\ke_{\tilde{s}} \leq s < \ke_{\tilde{s}+1}$
for every $s \in \{0,1,\ldots \ke_{I}-1\}$.}
}

The last result we will need from \cite{eymg} is the following
lemma.
\begin{lem} \label{projc}\mnote{projc}
If $X\in V_{2}$, $\ke_{\tilde{p}} + s < \ke_{\tilde{p}+1}$
$(s\geq 0)$, and $\Op^{\ke_{\tilde{p}}+s} . \, X = 0$, then
$\Op^{\ke_{\tilde{p}}}  . \, X = 0$.
\end{lem}

We will also frequently use the following fact
\leqn{lpm1}{
(l\pm 1)\tilde{\;} = \tilde{l} \pm 1 \quad \forall \: l\in \Ec \: . 
}
 
\begin{prop} \label{algthm11} \mnote{[algthm11]}
If $X_{a} \in E^{\tilde{a}}$, $Y_{b} \in E^{\tilde{b}}$
, and $Z_{c} \in E^{\tilde{c}}$
then $[[c(X_{a}),Y_{b}],Z_{c}] \in E^{(a+b+c)\tilde{\;}}$.
\end{prop}
\begin{proof}
Suppose $X_{a} \in E^{\tilde{a}}$, $Y_{b} \in E^{\tilde{b}}$, and
$Z_{c} \in E^{\tilde{c}}$. Then
\leqn{algthm11.1}{ 
\Op^{\ke_{\tilde{a}}+2}.c(X_{a}) = \Op^{\ke_{\tilde{b}}}.Y_{a}
= \Op^{\ke_{\tilde{c}}}.Z_{c} = 0
}

by lemma \ref{proja}. Now,
\eqn{algthm11.2}{
\Op^{p}.[[c(X_{a}),Y_{b}],Z_{c}]
= \sum_{l=0}^{p} \sum_{m=0}^{l} \binom{p}{l} \binom{l}{m}
W_{pabclm}
}
where $W_{pabclm} = [[\Op^{m}.c(X_{a}),\Op^{l-m}.Y_{b}],\Op^{p-l}.Z_{c}]$.
It then follows from \eqref{algthm11.1} that $W_{pabclm}=0$ if
$m\geq\ke_{\tilde{a}}+2$ or $l-m\geq\ke_{\tilde{b}}$ or $p-l\geq
\ke_{\tilde{c}}$. Thus $W_{pabclm}=0$ unless $p<l+\ke_{\tilde{c}}<m +
\ke_{\tilde{b}}+\ke_{\tilde{c}} <
\ke_{\tilde{a}}+\ke_{\tilde{b}}+\ke_{\tilde{c}}+2$.  But this can never be
satisfied if $p= \ke_{\tilde{a}}+\ke_{\tilde{b}}+\ke_{\tilde{c}}$ and so we
arrive at $\Op^{ \ke_{\tilde{a}}+\ke_{\tilde{b}}+\ke_{\tilde{c}}}
.[[c(X_{a}),Y_{b}],Z_{c}] = 0$. But
$\ke_{\tilde{a}}+\ke_{\tilde{b}}+\ke_{\tilde{c}} \leq a+b+c$ by
\eqref{tildeprop} and hence it follows that $\Op^{a+b+c}
.[[c(X_{a}),Y_{b}],Z_{c}] = 0$. But then lemma \ref{projc} implies that
$\Op^{\ke_{(a+b+c)\tilde{\;}}}.[[c(X_{a}),Y_{b}],Z_{c}] = 0$ and hence
$[[c(X_{a}),Y_{b}],Z_{c}] \in E^{(a+b+c)\tilde{\;}}$.
\end{proof}

\begin{prop} \label{algthm12} \mnote{[algthm12]}
If $X_{a} \in E^{\tilde{a}}$ and $Y_{b} \in E^{\tilde{b}}$ then 
$[[X_{a},c(Y_{b})],\Op]$, $[[\Op,X_{a}],Y_{b}]$, 
$[[\Om,X_{a}],Y_{b}]$  $\in E^{(a+b)\tilde{\;}}$.
\end{prop}
\begin{proof}
This proposition can be proved using the same techniques
as proposition \ref{algthm11}.
\end{proof}

\begin{lem} \label{algthm13} \mnote{[algthm13]}
If $l\in \Ec$ and $Z\in E^{\tilde{l}}_{+} \oplus E^{\tilde{l}-1}$
then $\Op^{l+1}.c(Z) = l(l+1)\Op^{l-1}.Z\:$.
\end{lem}
\begin{proof}
Since $Z\in E^{\tilde{l}}_{+} \oplus  E^{\tilde{l}-1}$
there exist real constants
$a_{q}^{s}$, $b_{q}^{s}$ such that 
\leqn{algthm13.2}{
Z = \sum_{q=1}^{\tilde{l}-1}\sum_{s=0}^{m_{q}-1}(a_{q}^{s}
+ i b^{s}_{q}) X^{q}_{s} + \sum_{s=0}^{m_{\tilde{l}}-1} 
a^{s}_{\tilde{l}} X^{\tilde{l}}_{s}
}
where $X^{q}_{s} = \xi^{J_{q}+s}_{\ke_{q}-1}$ if $\ke_{q}$ is odd
and $X^{q}_{s} = i\xi^{J_{q}+s}_{\ke_{q}-1}$ if $\ke_{q}$ is even. 
But
\leqn{algthm13.4}{ 
\Op^{l-1}.X^{q}_{s} = 0 \quad \text{for $q\leq\tilde{l}-1$} 
}
by lemma \ref{proja}, and so we get
\leqn{algthm13.5}{
\Op^{l-1}.Z = \sum_{s=0}^{m_{\tilde{l}}-1}
a^{s}_{\tilde{l}}\Op^{l-1}.X^{\tilde{l}}_{s} \;.
}
Now, $c(X^{q}_{s}) = \xi^{J_{q}+s}_{\ke_{q}+1} $ if $\ke_{q}$ is odd
and $c(X^{q}_{s}) = i\xi^{J_{q}+s}_{\ke_{q}+1}$ $\ke_{q}$ is even
by proposition \ref{hwv} and so 
\leqn{algthm13.7}{
\Op^{2}.c(X^{q}_{s}) = \ke_{q}(\ke_{q}+1) X^{q}_{s} \: .
}
Since $l\in \Ec$ implies that $\ke_{\tilde{l}}=l$, it follows easily form
\eqref{algthm13.2},\eqref{algthm13.4}, and \eqref{algthm13.7} that
\leqn{algthm13.8}{
\Op^{l+1}.c(Z) = l(l+1)\sum_{s=0}^{m_{\tilde{l}}-1}
a^{s}_{\tilde{l}}\Op^{l-1}.X^{\tilde{l}}_{s}\: .
}
Comparing \eqref{algthm13.5} and \eqref{algthm13.8}, we see that
$\Op^{l+1}.c(Z) = l(l+1)\Op^{l-1}.Z\:$.
\end{proof}

\begin{prop} \label{algthm14} \mnote{[algthm14]}
If $Z_{l}$ with $l=0,1,\ldots k$ is a sequence of vectors such that
\gath{algthm14.1}{
Z_{0} = \Op \: , \quad Z_{l} \in E^{\tilde{l}} \quad l=1,2,\ldots k
\AND
Z_{l} \in E^{l}_{+}\oplus E^{\tilde{l}-1} \quad \text{if $l\in \Ec$}
}
then for any $j=1,\ldots,l$
\eqn{algthm14.2}{
\sum_{s=1}^{k-1}[[c(Z_{j-s}),Z_{s}],\Op]
\in \begin{cases}
E^{\tilde{k}} & \text{if $k \notin \Ec$} \\
E^{\tilde{k}-1} & \text{if $k\in \Ec$}
\end{cases} \: .
}
\end{prop}

\begin{proof}
Suppose $Z_{l}$ is as in the hypotheses of the proposition, then
\leqn{algthm14.4}{
\Op^{\ke_{\tilde{l}}+2}.c(Z_{l}) = \Op^{\ke_{\tilde{l}}}.Z_{l} = 0 \: .
}
Now,
\leqn{algthm14.5}{
-\Om^{p}.\sum_{s=1}^{k-1} [[c(Z_{k-s},Z_{s}],\Op]
= \sum_{s=1}^{k-1} \sum_{l=0}^{p+1} \binom{p+1}{l}
W_{pkls}
}
where $W_{pkls} = [\Op^{l}.c(Z_{k-s}),\Op^{p+1-l}.Z_{s}]$.
From \eqref{algthm14.4} we see that $W_{pkls} = 0$ if $l\geq
\ke_{(k-s)\tilde{\;}}$ or $p+1-l\geq \ke_{\tilde{s}}$.  Thus
\leqn{algthm14.7}{
\text{$W_{pkls}=0$ unless $p+1 < l+\ke_{\tilde{s}}
< \ke_{\tilde{s}} + \ke_{(k-s)\tilde{\;}} +2$} \;.
}

Now, suppose $p=k$. Then using the fact that $\ke_{\tilde{s}} \leq s$ and
$\ke_{(k-s)\tilde{\;}} \leq k-s$, we get from \eqref{algthm14.7} that
$W_{pkls} = 0$ unless $k+1 < l+\ke_{\tilde{s}}< k+2$. Since this is impossible
to satisfy $W_{pkls} = 0$ for all $l,s$.  Thus the sum \eqref{algthm14.5}
vanishes, i.e. $ -\Om^{k}.\sum_{s=1}^{k-1} [[c(Z_{k-s},Z_{s}],\Op] = 0$,
and we get
\leqn{algthm14.9}{
\sum_{s=1}^{k-1} [[c(Z_{k-s},Z_{s}],\Op] \in
E^{\tilde{k}}
}
by lemmas \ref{proja} and \ref{projc}. Suppose 
further that $k\in\Ec$ and let $p=k-1$. Then 
$a_{pkls}=0$ unless $k < l+\ke_{\tilde{s}} <
\ke_{\tilde{s}} + \ke_{(k-s)\tilde{\;}} +2$ by \eqref{algthm14.7}.  
Now, $\ke_{\tilde{s}} \leq s$ and
$\ke_{(k-s)\tilde{\;}} \leq k-s$, so suppose $\ke_{\tilde{s}} < s$ or
$\ke_{(k-s)\tilde{\;}} < k-s$. Then $\ke_{(k-s)\tilde{\;}} + \ke_{\tilde{s}} <
k-s+s=k$ which will make the inequality $k < l+\ke_{\tilde{s}} <
\ke_{\tilde{s}} + \ke_{(k-s)\tilde{\;}} +2$ impossible to satisfy. Therefore
we see that $W_{pkls}=0$ unless $ \ke_{(k-s)\tilde{\;}} = k-s$ and
$\ke_{\tilde{s}}=s$ (i.e. $k-s,s \in \Ec$). However, if $
\ke_{(k-s)\tilde{\;}} = k-s$ and $\ke_{\tilde{s}}=s$, then $k <
l+\ke_{\tilde{s}} < \ke_{\tilde{s}} + \ke_{(k-s)\tilde{\;}} +2$ will satisfied
only if $l+s=1$. So $W_{pkls}=0$ unless $k-s,s\in\Ec$ and $l+s=1$.  This
allows us to write the sum \eqref{algthm14.5} as
\alin{algthm14.10}{
-\Om^{k-1}.  \sum_{s=1}^{k-1}& [[c(Z_{k-s}),Z_{s}],\Op] \\
=& \sum_{s=1}^{k-1} \binom{k}{k-s+1}(k-s)(k-s+1)
[\Op^{k-s-1}.Z_{k-s},\Op^{s-1}.Z_{s}] && \text{by lemma \ref{algthm12}} \\
=& \sum_{s=1}^{k-1}\frac{k!}{(k-s-1)!(s-1)!}
[\Op^{k-s-1}.Z_{k-s},\Op^{s-1}.Z_{s}] \: .
}

Assume now that $k$ is odd. Then we can write the above sum as
\gath{algthm14.11}{
-\Om^{k-1}.\sum_{s=1}^{k-1} [[c(Z_{k-s}),Z_{s}],\Op]
= \sum_{s=\frac{k-1}{2}-1}^{k-1}\frac{k!}{(k-s-1)!(s-1)!}
[\Op^{k-s-1}.Z_{k-s},\Op^{s-1}.Z_{s}] \\ 
 + \sum_{s=1}^{\frac{k-1}{2}}\frac{k!}{(k-s-1)!(s-1)!}
[\Op^{k-s-1}.Z_{k-s},\Op^{s-1}.Z_{s}]  \\
 = \sum_{s=1}^{\frac{k-1}{2}}\frac{k!}{(k-s-1)!(s-1)!}\left\{
[\Op^{k-s-1}.Z_{k-s},\Op^{s-1}.Z_{s}] +
[\Op^{s-1}.Z_{s},\Op^{k-s-1}.Z_{k-s}]\right\} = 0 \; .
}

Similar arguments show that $-\Om^{k-1}.\sum_{s=1}^{k-1} 
[[c(Z_{k-s},Z_{s}],\Op]=0$ if $k$ is even. Therefore
$\sum_{s=1}^{k-1}[[c(Z_{j-s}),Z_{s}],\Op]
\in E^{\tilde{k}-1}$
by lemmas \ref{proja} and \ref{projc} and \eqref{lpm1}.
\end{proof}

\begin{prop} \label{algthm15} \mnote{[algthm15]}
If $Z_{l}$ with $l=0,1,\ldots k$ is a sequence of vectors such that
\gath{algthm15.1}{
Z_{0} = \Op \: , \quad Z_{l} \in E^{\tilde{l}} \quad l=1,2,\ldots k 
\AND
Z_{l} \in E^{l}_{+}\oplus E^{\tilde{l}-1}\quad \text{if $l\in \Ec$}
}
then, for any $j=1,\ldots,l$
\eqn{algthm15.2}{
\sum_{j=1}^{k-1}\sum_{s=0}^{j}[[c(Z_{j-s}),Z_{s}],Z_{k-j}]
\in \begin{cases}
E^{\tilde{k}} & \text{if $k\notin \Ec$} \\
E^{\tilde{k}-1} & \text{if $k\in \Ec$}
\end{cases} \: .
}
\end{prop}
\begin{proof}
Proved using similar arguments as for proposition \ref{algthm14}.
\end{proof}

\sect{lep}{Local Existence Proofs}

For $q=1,2,\ldots I\:$ let
\leqn{eproj}{
\pr^{q}_{+} : V_{2} \rightarrow E^{q}_{+}\:, \;\; 
\pr^{q}_{0} : V_{2} \rightarrow E^{q}_{0}\: ,
\AND 
\pr^{q} : V_{2} \rightarrow E^{q}_{0}\oplus E^{q}_{+}
}
denote the projections determined by the decomposition \eqref{V2},
\eqref{E0} of $V_{2}$.

\subsect{origin}{Solutions regular at the origin}

\begin{thm} \label{origthm1} \mnote{[origthm1]}
  Fix $X\in E_{+}$ and $\Op \in E_{+}$ that satisfies $[\Op,\Om] = \Lo$ where
  $\Om := -c(\Op)$.  Then there exist a unique solution $\{\Lp(r,Y), m(r,Y)\}$
  to the system of differential equations \eqref{feq1} and \eqref{feq3} that
  is analytic in a neighborhood of $(r,Y)=(0,X)$ in $\mathbb{R}\times E_{+}$
  and satisfies $m = \text{O}(r^{3})$ and
\eqn{orightm1.1a}{
\emph{\pr}^{\tilde{s}}_{+}(\Lp-\Op) = Y_{s} r^{s+1} + \text{O}(r^{s+2})\:, \quad
\emph{\pr}^{\tilde{s}}_{0}(\Lp) = \text{O}(r^{s+2}) \quad \forall s\in \Ec
}

  where $\Op := \Lp(0)$ and $Y_{s} := \emph{\pr}^{\tilde{s}}_{+}(Y)$.  
Moreover, these
solutions also satisfy $P=\text{O}(r^{4})$ and $G=\text{O}(r^{2})$.
\end{thm}
\begin{proof}
Introduce new variables
$\{\:u^{+}_{s},\: u^{0}_{s} \: | \: s\in \Ec\:\}$ via
\leqn{orightm1.1b}{
u^{+}_{s} := \pr^{\tilde{s}}_{+}(\Lp-\Op)r^{-s-1} \AND
u^{0}_{s} := \pr^{\tilde{s}}_{0}(\Lp)r^{-s-2} 
}

where $\Op := \Lp(0)$. This allows us to write $\Lp$ as 
\leqn{origthm1.1}{
\Lp(r) = \Op  + \sum_{s\in \Ec} ( u^{+}_{s}(r) + r u^{0}_{s}(r))\, r^{s+1}
\: .
}

But %It is important to realize that 
regularity at $r=0$ requires that $\Op$ satisfy $[\Op,\Om] = \Lo$ where $\Om
:= -c(\Op)$.  This can be seen easily from \eqref{vardefs2}, \eqref{vardefsa},
\eqref{ener} and the requirement that the pressure remains finite at $r=0$.

\begin{lem} \label{origthm2} \mnote{[orgthm2]}
For every $s\in \Ec$ there  exists analytic maps
$\Fc^{1}_{s} : E_{+} \rightarrow E^{\tilde{s}}_{0}
\oplus E^{\tilde{s}}_{+}$ and 
$\Fc^{2}_{s} : E_{0}\times E_{+}\times \mathbb{R}
 \rightarrow E^{\tilde{s}}_{0}
\oplus E^{\tilde{s}}_{+}$
such that, for $\Fc$ given in \eqref{vardefs3},
$\emph{\pr}^{\tilde{s}}\Fc = -s(s+1) u^{+}_{s} r^{s+1} + 
r^{s+2}\Fc^{1}_{s}(u^{+}) + r^{s+3}\Fc^{2}_{s}(u^{0},u^{+},r)$
where
\leqn{origthm2.3}{
u^{0} := \sum_{a\in \Ec} u^{0}_{a} \AND
u^{+} := \sum_{a\in \Ec} u^{+}_{a}\: .
}
\end{lem}
\begin{proof}
Let $u_{s} =  ru^{0}_{s} + u^{+}_{s}$. Then from \eqref{vardefs3}
we find
\gath{origthm2.4}{
\Fc = \sum_{a\in \Ec} A_{2}(u_{a}) r^{a+1} + \frac{1}{2}{\DS\sum_{a,b\in \Ec}}
\left( [[u_{a},c(u_{b})],\Op]+[[\Op,c(u_{a})],u_{b}] +
[[\Om,u_{a}],u_{b}] \right)r^{a+b+2} \\
+ \frac{1}{2}{\DS\sum_{a,b,c\in \Ec}} [[u_{a},c(u_{b})],u_{c}] r^{a+b+c+3} \: .
}

But from \eqref{A2x} we see that 
\leqn{origthm2.5}{
A_{2}(u_{a})= a(a+1) u^{+}_{a} \: .
} 

Also, for $a\in \Ec$ we have $\ke_{\tilde{a}}=a$ by lemma \ref{tildeprop},
and hence
\leqn{origthm2.6}{
\tilde{a} \leq \tilde{b} \Longleftrightarrow a\leq b\: .
}

Using the two results \eqref{origthm2.5} and \eqref{origthm2.6}, we get
\gath{origthm2.7}{
\pr^{\tilde{s}}\Fc = -s(s+1) u^{+}_{0} r^{s+1}
 + \frac{1}{2}{\DS\sum_{\substack{a,b\in \Ec\\ a+b\geq s}}} 
\pr^{\tilde{s}}\biggl( [[u_{a},c(u_{b})],\Op]+
[[\Op,c(u_{a})],u_{b}] \biggr.\\ 
+\biggl. [[\Om,u_{a}],u_{b}] \biggr)r^{a+b+2} 
+\frac{1}{2}{\DS\sum_{\substack{a,b,c\in \Ec\\ a+b+c\geq s}}} 
\pr^{\tilde{s}}\biggl([[u_{a},c(u_{b})],u_{c}]\biggr) r^{a+b+c+3} 
}

by propositions \ref{algthm11} and \ref{algthm12}.
Substituting  $u_{a} =  ru^{0}_{a} + u^{+}_{a}$ into the
above expression completes the proof.
\end{proof}
For every $s\in \Ec$ define
\leqn{origthm1.3}{
v^{+}_{s} := {u_{s}^{+}}' \AND v^{0}_{s} := (r u^{0}_{s})' \: .
}

\begin{lem} \label{origthm3} \mnote{[orgthm3]}
There exists analytic functions
$\hat{P} : E_{0}\times E_{+} \times \mathbb{R} \rightarrow \mathbb{R} $
and $\hat{G} : E_{0}\times  E_{0}\times E_{+} \times E_{+} \times 
\mathbb{R} \rightarrow \mathbb{R}$
such that 
$P=r^{4}\norm{u^{+}_{1}}^{2} + r^{5}\hat{P}(u^{0},u^{+},r)$
and $G = r^{2} \norm{u^{+}_{1}}^{2} + r^{3}
\hat{G}(u^{0},v^{0},u^{+},v^{+},r)$ where
\leqn{origthm3.3}{
v^{0} := \sum_{a\in \Ec} v^{0}_{a} \AND
v^{+} := \sum_{a\in \Ec} v^{+}_{a}\: .
}
and $u^{0},\; u^{+}$ are defined by \eqref{origthm2.3}.
\end{lem}
\begin{proof}
The existence of the analytic function $\hat{G}$ follows easily from the
definition \eqref{vardefs1} of $G$ and equations \eqref{origthm2.3} and
\eqref{origthm3.3}.  

Similarly from the definition \eqref{vardefs1} of $P$ 
we have %it is not hard to see that
$P = \frac{r^{4}}{8}\norm{[\Op,c(u_{1}^{+})]+[\Om,u_{1}^{+}]}
 + r^{5}Q(u^{0},u^{+},r)$
where $Q$ is a polynomial in $r$, $u^{0}$ and $u^{+}$.  Now, using
\eqref{rip}, \eqref{Adef}, and $A_{2}(u^{+}_{1})=2u^{+}_{1}$, it
is not difficult to show that
$\frac{1}{8}\norm{[\Op,c(u_{1}^{+})]+[\Om,u_{1}^{+}]}
= \norm{u^{+}_{1}}$
and this completes the proof.
\end{proof}

From \eqref{El}-\eqref{kerA2} we get
$A_{2}(u^{+}_{s}) = s(s+1)u^{+}_{s}$ and $A_{2}(u^{0}_{s})=0$
since $u^{+}_{s} \in E^{\tilde{s}}_{+}$ and $u^{0}_{s} \in E^{\tilde{s}}_{0}$.
Using this result, lemma \ref{origthm2}, and equations \eqref{origthm2.3} and
\eqref{origthm3.3}, the field equations \eqref{feq1} and \eqref{feq3} can be
written as
\lalign{origthm1.5}{
r{u^{+}_{s}}' & = rv^{+}_{s} \: , \label{origthm1.5.1}\\
r{v^{+}_{s}}' & = -2 (s+1) v^{+}_{s} - \frac{2}{r N}\left(m-\frac{1}{r} 
P\right)v^{+}_{s}  - \frac{s(s+1)}{r} \left(\frac{1}{N}-1\right) 
u^{+}_{s} \notag \\
& -\frac{2(s+1)}{r^{2} N}\left(m-\frac{1}{r} P\right)u^{+}_{s}
 -\frac{r}{N}\pr^{\tilde{s}}_{+}\Fc^{2}_{s}(u^{0},u^{+},r) \notag \\
&  -\left(\frac{1}{N}-1\right)\pr^{\tilde{s}}_{+}\Fc^{1}_{s}(u^{+})
-\pr^{\tilde{s}}_{+}\Fc^{1}_{s}(u^{+}) \: , \label{origthm1.5.2} \\
r{u^{0}_{s}} ' &= - u^{0}_{s} + v^{0}_{s} \; , \label{origthm1.5.3} \\
r{v^{0}_{s}}' & = -2 (s+1) v^{0}_{s} - s(s+1)u^{0}_{s}
-\frac{2}{N}\left(m-\frac{1}{r} 
P\right)v^{0}_{s} \notag \\
& -\frac{2(s+1)}{rN}\left(m-\frac{1}{r} P\right)u^{0}_{s}
 -\frac{r}{N}\pr^{\tilde{s}}_{0}\Fc^{2}_{s}(u^{0},u^{+},r) \notag \\
&  -\left(\frac{1}{N}-1\right)\pr^{\tilde{s}}_{0}\Fc^{1}_{s}(u^{+})
-\pr^{\tilde{s}}_{0}\Fc^{1}_{s}(u^{+}) \; , \label{origthm1.5.4} 
}
where $s\in \Ec$.  For every $s\in \Ec$, introduce two new variables
\eqn{origthm1.6}{
x_{s} := -(s+1)u^{0}_{s} - \frac{s+1}{s}v^{0}_{s} \:, \quad
y_{s} := (s+1)u^{0}_{s} + v^{0}_{s}\: , 
}
and define
\alin{origthm1.7}{
f_{s}  :=  -\frac{2}{N}&\left(m-\frac{1}{r}
P\right)v^{0}_{s} 
 -\frac{2(s+1)}{rN}\left(m-\frac{1}{r} P\right)u^{0}_{s} \\
& -\frac{r}{N}\pr^{\tilde{s}}_{0}\Fc^{2}_{s}(u^{0},u^{+},r)
  -\left(\frac{1}{N}-1\right)\pr^{\tilde{s}}_{0}\Fc^{1}_{s}(u^{+})
\; . 
}
Then equations \eqref{origthm1.5.3} and \eqref{origthm1.5.4} can be written as
\lalign{origthm1.8}{
rx_{s}' &= -(s+2) x_{s} + \frac{s+1}{s}\pr^{\tilde{s}}_{0}\Fc^{1}_{s}(u^{+})
+\frac{s+1}{s} f_{s}  ,\label{origthm1.8.1} \\
ry_{s}' &= -(s+2) y_{s} -\pr^{\tilde{s}}_{0}\Fc^{1}_{s}(u^{+})
f_{s} \: , \label{origthm1.8.2} 
}
for every $s\in \Ec$. Define 
$\mu = r^{-3}\left(m-r^{3}\norm{u^{+}_{1}}^{2}\right)$.
Then the mass equation \eqref{feq1} becomes
\lalign{origthm1.10}{
r \mu' = -3 \mu + r&\left\{  \hat{P}(u^{0},u^{+},r) + 
\hat{G}(u^{0},v^{0},u^{+},v^{+},r) -
        2 \rip{u^{+}_{1}}{v^{+}_{1}} \right. \nonumber \\
        & \left. - 2\, r\, \left(\mu + \norm{u^{+}_{1}}^{2}\right)
        \left( 2 \norm{u^{+}_{1}}^{2} + r \hat{G}(u^{0},v^{0},u^{+},v^{+},r) 
\right) \
\right\} \; .  \label{origthm1.10.2}
}
For every $s\in \Ec$, introduce one last change of variables
\eqn{origthm1.11a}{
\hat{v}^{+}_{s} := v^{+}_{s} + \frac{1}{2(s+1)}
\pr^{\tilde{s}}_{0}\Fc^{1}_{s}(u^{+}) \: , \quad
\hat{x}_{s} :=  x_{s} - \frac{s+1}{s(s+2)}
\pr^{\tilde{s}}_{0}\Fc^{1}_{s}(u^{+})  \:, \AND \hat{y}_{s}  :=  
y_{s} + \frac{1}{(s+1)}\pr^{\tilde{s}}_{0}\Fc^{1}_{s}(u^{+})\: .
}
Let $\hat{v}^{+} :=  \sum_{s\in \Ec} v^{+}_{s}$, $\hat{x} :=  
\sum_{s\in \Ec} x_{s}$, $\hat{y} :=   \sum_{s\in \Ec} y_{s}$ 
and
$\eta(r) := (\hat{x}(r),\hat{y}(r),u^{+}(r),\hat{v}^{+}(r),\mu(r),r)$.
Fix $X\in E_{+}$ and let $\Nc_{X}$ be a neighborhood of $X$ in $E_{+}$.
Define a set $D(\Nc_{X},\epsilon)$ by
$D(\Nc_{X},\epsilon) := E_{0}\times E_{0} \times \Nc_{X} \times
E^{+} \times (-\epsilon, \epsilon) \times (-\epsilon, \epsilon)$.
Then using lemmas \ref{origthm2} and \ref{origthm3} and equations
\eqref{origthm1.5.1}, \eqref{origthm1.5.2}, \eqref{origthm1.8.1},
\eqref{origthm1.8.2}, and \eqref{origthm1.10.2} one can show that there exists
an $\epsilon > 0$ and analytic maps 
$\Uc_{s} , \Vc_{s} : D(\Nc_{X},\epsilon) \rightarrow E^{\tilde{s}}_{+}$
, $\Xc_{s} , \Yc_{s} : D(\Nc_{X},\epsilon) \rightarrow E^{\tilde{s}}_{0}$
, and  $\Mc : D(\Nc_{X},\epsilon) \rightarrow \mathbb{R}$ 
such that for all $s\in \Ec$
\gath{orightm1.16}{ r{u_{s}^{+}} '(r)
  = r\Uc_{s}(\eta(r)), \quad {r\hat{v}_{s}^{+}} ' (r) =
  -2(s+1)\hat{v}_{s}^{+}(r) +
  r\Vc_{s}(\eta(r)), \notag  \\
  {r\hat{x}_{s}} '(r) = -(s+2)\hat{x}_{s}(r) + r\Xc_{s}(\eta(r)) , \quad
  {r\hat{y}_{s}} '(r) = -(s+1)\hat{x}_{s}(r) + r\Yc_{s}(\eta(r)) , \notag 
}
and ${r\mu} '(r) = -3\mu(r) + r\Mc(\eta(r))$.
This system of differential equations is in the form to which theorem
\ref{BFM} applies. Applying this theorem shows that for fixed $X \in E_{+}$
there exist a unique solution $\{u^{+}_{s}(r,Y)$,$\hat{v}^{+}_{s}(r,Y)$,
$\hat{x}_{s}(r,Y)$,$\hat{v}_{s}(r,Y),\mu(r,Y)\}$ that is analytic in a
neighborhood of $(r,Y) = (0,X)$ and that satisfies
$u^{+}_{s}(r,Y) = Y_{s} + \text{O}(r)$, $\hat{v}^{+}_{s}(r,Y) =
  \text{O}(r)$, $ \hat{x}_{s}(r,Y) = \text{O}(r)$, $\hat{y}_{s}(r,Y) =
  \text{O}(r)$, and  $\mu(s,Y) = \text{O}(r)$
where $Y_{s} = \pr^{\tilde{s}}_{+}(Y)$. From these
results it is not hard to verify that 
$m(r) = \text{O}(r^{3})$ and $ u^{0}(r) = \text{O}(r^{0}) $
Also, it is clear that
$P= \text{O}(r^{4})$ and  $G = \text{O}(r^{2})$ 
by lemma \ref{origthm3}.
\end{proof}

\begin{thm} \label{origthm4} \mnote{[orightm4]}
Every solution from theorem \ref{origthm1} satisfies
equation \eqref{feq4} on a neighborhood of $r=0$.
\end{thm}
\begin{proof}
Let $\{\Lp(r),m(r)\}$ be a solution of the equations
\eqref{feq1} and \eqref{feq3} on a neighborhood $\Nc$ of $r=0$, which
we know exists by theorem \ref{origthm1}.
From theorem \ref{origthm1} it is clear that $\Lp'(0)=0$, and this implies that
$\gamma(0)=0$ where $\gamma(r)$ is defined in lemma \ref{constraint}.
Also, because $m=\text{O}(r^{3})$ and $P=\text{O}(r^{4})$ for 
these solutions we see, by shrinking $\Nc$ if necessary, that the function 
$f(r) = -2(r^2N)^{-1}\left(m-\frac{1}{r}P\right)$
is analytic on $\Nc$. From lemma \ref{constraint}, $\gamma$ satisfies
the differential equation
$\gamma' = f(r)\gamma $.
Solving this equation we find
$\gamma(r) = \gamma(0) \exp(\int_{0}^{r} f(\tau) d\tau)$  for all
$r \in \Nc$ .
But $\gamma(0)=0$, so $\gamma(r)=0$ for all $r\in \Nc$.
\end{proof}

\subsect{infty}{Asymptotically flat solutions}

In proving that local solutions exist near $r=0$, we were able
to ``guess'' the appropriate transformations needed to bring
the field equations \eqref{feq1} and \eqref{feq3} in to a
form for which theorem \ref{BFM} applies. Near, $r=\infty$ the
equations become much more difficult to analyze and guessing
the appropriate transformation is no longer possible.
Instead, we will show that the field equations 
\eqref{feq1} and \eqref{feq3} admit a formal power series solutions
about the point $r=\infty$. This formal power series solution will
then be used to construct a transformation to bring the
equations \eqref{feq1} and \eqref{feq3} in to a
form for which theorem \ref{BFM} applies. 

Let $z = \frac{1}{r}$, and define
\eqn{zdot}{
\dz{f} = \frac{df}{dz} 
}
for any function $f$. Then the field equations \eqref{feq1} and \eqref{feq3}
can be written as
\lgath{zfeq}{
z^{2}\dz{m} + NG + z^{2}P = 0\: , \label{zfeq1}\\
z^{2}N {\dzt{\Lambda}}_{+} + 2z(1-3zm+z^{2}P){\dz{\Lambda}}_{+} + \Fc=0\:.\label{zfeq3}
}
Assume a powerseries expansion of the form
\leqn{power}{
\Lp = \sum_{k=0}^{\infty} \Lpp{k} z^{k} \AND
m = \sum_{k=0}^{\infty} m_{k} z^{k} \; .
}
We will define
$\Lmm{k} := -c(\Lpp{k})$ and $\Opm := \La_{\pm,0}$.
From the requirement that the total magnetic charge vanish we have that
$[\Op,\Om] = \Lo$.  Substituting the powerseries \eqref{power} in the
equations \eqref{zfeq1} and \eqref{zfeq3} yields the recurrence equations
\lgath{recur}{
m_{1}=m_{2}=0\:, \quad m_{k} = \frac{1}{k}
\left(-G_{k-3}+2\sum^{k-4}_{j=0}m_{j} G_{k-j-4} - P_{k-1} \right)
\quad k=3,4,5,\ldots \label{recur1} \\
A_{2}(\Lpp{k})-k(k+1)\Lpp{k} = h_{k} + f_{k} \quad k=1,2,3,\ldots
\label{recur2}
}
where
\lgath{varsinfty}{
G_{k} := \frac{1}{2} \sum_{j=0}^{k} (j+1)(k+1-j)\rip{\Lpp{k+1-j}}{\Lpp{j+1}}
\quad k \geq 0\: , \label{varsinfty1} \\
\Fh_{0}:=0\:, \quad \Fh_{k} := \frac{1}{2} \sum_{j=1}^{k}
\sum_{s=0}^{j} [[\Lmm{j-s},\Lpp{s}],\Lpp{k-j}] \quad k\geq 1\: ,
\label{varsinfty2} \\
P_{0}=P_{1}=0,\quad P_{k} := \frac{1}{2}\sum_{j=1}^{k-1}\rip{\Fh_{j}}{\Fh_{k-j}} \quad 
k \geq 2\: ,
\label{varsinfty3} \\
h_{1} = 0 , \quad h_{k} := 2\sum_{j=0}^{k-1} (k-j-1)\left(P_{j-2} - (k-j+1)m_{j}
\right) \Lpp{k-j-1} \quad k \geq 2\: ,
\label{varsinfty4} \\
\intertext{and}
f_{1} = 0 , \quad f_{k} := \frac{1}{2}\left\{
\sum_{j=1}^{k-1} \sum_{s=0}^{j} [[\Lmm{j-s},\Lpp{s}],\Lpp{k-j}]
+ \sum_{s=1}^{k-1}[[\Lmm{k-s},\Lp{s}],\Op]
\right\} \quad k\geq 2\: .  
\label{varsinfty5}
}
Note that with these definitions that
$\Fh = \sum_{k=0}^{\infty}\Fh_{k}z^{k}$ and 
$P = \sum_{k=0}^{\infty}P_{k}z^{k}$ while 
$G =  \sum_{k=0}^{\infty} G_{k}z^{k+4}$.

\begin{thm} \label{inftythm1} \mnote{[inftythm1]}
Fix $X \in E_{+}$ and $m_{\infty}\in \mathbb{R}$. 
Then there exists a unique solution
$\{\Lpp{k},m_{k}\}_{k=0}^{\infty}$ to the recurrence 
equations \eqref{recur1} and \eqref{recur2} that satisfies
\lgath{inftythm1.1}{
 m_{0} = m_{\infty}\:, \quad m_{1}=m_{2}=0  \quad ; \quad
\emph{\pr}^{\tilde{k}}_{+}\Lpp{k} = X_{k} \quad  \forall\; k\in \Ec \\
\intertext{and}
\Lpp{k} \in \begin{cases}
E^{\tilde{k}} & \text{if $k\notin \Ec$}\\
 E^{\tilde{k}}_{+}\oplus E^{\tilde{k}-1}& \text{if $k\in \Ec$} 
\end{cases} \: ,
}
where $X_{k} :=  \emph{\pr}^{\tilde{k}}_{+} X$.
\end{thm}
\begin{proof}
Fix $X\in E_{+}$, $m_{\infty}\in \mathbb{R}$, and let
 $ X_{k} =  \pr^{\tilde{k}}_{+} X$ for all $k\in \Ec$. 
We will use induction to prove that the recurrence equations 
\eqref{recur1} and \eqref{recur2} can be solved. When $k=1$, the
equations \eqref{recur1} and \eqref{recur2} reduce to
$m_{1} = 0$ and  $A_{2}(\Lpp{1}) - 2\Lpp{1} = 0$.
This can be solved in $E^{1}_{0}\oplus E^{1}_{+}$ by letting
$\Lpp{1} = X_{1}$. Note that since $\min{\Ec}=1$, we have
$\tilde{1}=1$. 

We now assume that for $k\leq l$ , $\{\Lpp{k},m_{k}\}$ is
a solution to the recurrence equations \eqref{recur1} and \eqref{recur2}
that satisfies
\eqn{inftythm1.4}{
\Lpp{k} \in \begin{cases}
E^{\tilde{k}} & \text{if $k\notin \Ec$}\\
 E^{\tilde{k}}_{+}\oplus E^{\tilde{k}-1} & \text{if $k\in \Ec$} 
\end{cases} \: .
}

It is clear from \eqref{recur2} that $m_{l+1}$ is then determined.
From \eqref{varsinfty1}-\eqref{varsinfty5} and propositions
\ref{algthm14} and \ref{algthm15} it follows that
\leqn{inftythm1.5}{
h_{l+1}+f_{l+1} 
\in \begin{cases}
E^{\tilde{l}+1} & \text{if $l+1\notin  \Ec$} \\
E^{\tilde{l}} & \text{if $l+1\in \Ec$}
\end{cases} \: .
}

Equation \eqref{recur2} implies that
\leqn{inftythm1.6}{
\bigl(A_{2}-(l+1)(l+2)\id_{V_{2}}\bigr)\Lpp{l+1} = h_{l+1}+f_{l+1} \: .
}

Suppose $l+1 \notin \Ec$. Then $A_{2}-(l+1)(l+2)\id$ is invertible and
\eqn{inftythm1.7}{
\Lpp{l+1} = \bigl(A_{2}-(l+1)(l+2)\id\bigr)^{-1}( h_{l+1}+f_{l+1}) \: .
}

But then \eqref{inftythm1.5} implies that 
$\Lpp{l+1}\in E^{\tilde{l}+1}$.

Alternatively, suppose $l+1\in \Ec$.
Then 
$\ker\bigl(A_{2}-(l+1)(l+2)\id\bigr)
= E^{\tilde{l}+1}_{+}$ by \eqref{El}-\eqref{kerA2} and \eqref{lpm1}. 
Therefore, \eqref{inftythm1.5} shows that
\eqn{inftythm1.8}{
\Lpp{l+1} = \bigl(\left(A_{2}-(l+1)(l+2)\id\right)\restr{E^{\tilde{l}}}\bigr)^{-1}
\left( h_{l+1}+f_{l+1}\right) + X_{l+1}
}

solves \eqref{inftythm1.6} since $X_{l+1} \in \ker\bigl(A_{2}-(l+1)(l+2)\id\bigr)$. 
It also clear from \eqref{inftythm1.5} and $X_{l+1} \in E^{\tilde{l}+1}_{+}$
that  $\Lpp{l+1} \in E^{\tilde{l}+1} 
\bigoplus_{q=1}^{\tilde{l}} E^{q}_{0}\oplus E^{q}_{+}$. 
This prove that $\Lpp{l+1}$ satisfies the induction hypothesis and 
so the proof is complete.
\end{proof}

\begin{thm} \label{inftythm2} \mnote{[inftythm2]}
Fix $X \in E_{+}$, $m_{\infty} > 0$, and
$\Op \in E_{+}$ that satisfies
$[\Op,\Om] = \Lo$ where $\Om := -c(\Op)$. Then there exist a unique solution
$\{\Lp(r,a,Y)$, $m(r,a,Y)\}$ to the system of differential
equations \eqref{feq1} and \eqref{feq3} that is
analytic in the variables $(r^{-1},a,Y)$ in a neighborhood of 
$(0,m_{\infty},X) \in \mathbb{R}^{2} \times E_{+}$ and satisfies 
$m = a + \text{O}(r^{-3})$ and
\eqn{inftythm2.1}{
\emph{\pr}^{\tilde{s}}_{+}(\Lp-\Op) = \frac{Y_{s}}{r^{s}}+ 
\text{O}(r^{-s})\:, \quad
\emph{\pr}^{\tilde{s}}_{0}(\Lp) = 
\text{O}(r^{-s}) \quad \forall s\in \Ec
}
where $\Op := \Lp(0)$ and  $Y_{s} := \emph{\pr}^{\tilde{s}}_{+}(Y)$.
\end{thm}
\begin{proof}
Fix $X\in E_{+}$ and let 
$\Lpp{k} = \Lpp{k}(X,m_{\infty})$ and $m_{k} = m_{k}(X,m_{\infty})$
be solutions to the recurrence equations \eqref{recur1} and \eqref{recur2}
which satisfy
$ m_{0} = m_{\infty}$ , $m_{1}=m_{2}=0$,
$\pr^{\tilde{k}}_{+}\Lpp{k} = X_{k}$ for all $k\in \Ec$, and 
\eqn{inftythm2.3}{
\Lpp{k} \in \begin{cases}
\bigoplus_{q=1}^{\tilde{k}} E^{q}_{0}\oplus E^{q}_{+}
 & \text{if $k \notin \Ec$}\\
 E^{\tilde{k}}_{+}\oplus\bigoplus_{q=1}^{\tilde{k}-1}
E^{q}_{0}\oplus E^{q}_{+} & \text{if $k\in \Ec$}
\end{cases} \: ,
}
where $X_{k} = \pr^{\tilde{k}}_{+} X$.
Define
$U := \sum_{k=0}^{n} \Lpp{k}z^{k}$ and $M:= \sum_{k=0}^{n} m_{k}z^{k}$
and introduce new variables $\phi(z)$ and $\sigma(z)$ via 
\leqn{inftythm2.5}{
\Lp = U + z^{n-3}\phi \AND m = M + z^{n-3}\sigma \: ,
}
where the integer $n$ is to be chosen later.
Define
\lgath{inftythm2.6}{
N_{p} := 1-2Mz\: , \quad \Fh_{p} := \ihalf(\Lo+[U,c(U)])\: , 
\label{inftythm2.6.1} \\
\Fc_{p} := -i[\Fh_{p},U]\: , \quad P_{p} := \half\|Fh_{p}\|^{2}\: ,
\AND G_{p} := \half z^{4}\|\dz{U}\|^{2} \: .
\label{inftythm2.6.2}
}
From these definitions it is clear that the quantities $N_{p}$, $\Fh_{p}$,
$\Fc_{p}$,  $P_{p}$ and $G_{p}$ are all polynomial in the variables
$X$ and $m_{\infty}$.
Now,  because $U$ and $M$ are the first 
$n$ terms in the powerseries
solution to the field equations \eqref{zfeq1} and \eqref{zfeq3} about
the point $z=0$ they satisfy
\gath{inftythm2.7}{
z^{2}N_{p} \dzt{U} + 2z(1-3zM+z^{2} P_{p})\dz{U} + \Fc_{p}
= z^{n-1}(a_{1}(X,m_{\infty}) + a_{2}(X,m_{\infty})y) \: ,
\label{inftythm2.7.1} \\
z^{2}\dz{M} + N_{p}G_{p}+z^{2} P_{p} = z^{n} b(X,m_{\infty}) \: ,
\label{inftythm2.7.2}
}
where $a_{1}, a_{2} : V_{2}\times \mathbb{R} \rightarrow \mathbb{R}$ and
$b :  V_{2}\times \mathbb{R} \rightarrow \mathbb{R}$ are polynomial 
in their variables.

From \eqref{inftythm2.6.2} and \eqref{vardefs3} it
follows that
$\Fc  = \Fc_{p} - z^{n-3}A_{2}(\phi) + 
 z^{n-2}\sum_{j=1}^{3} \Fc_{R,j}(\phi,X,m_{\infty},z)$ 
where $\Fc_{R,j} : V_{2}\times E_{+} \times \mathbb{R} \times \mathbb{R}
\rightarrow \mathbb{R} \quad j=1,2,3$
are analytic maps that satisfy
$\Fc_{R,j}(\epsilon Y_{1},Y_{2},x_{1},x_{2}) =
\epsilon^{j}\Fc_{R,j}( Y_{1},Y_{2},x_{1},x_{2})$
for all $\epsilon \in \mathbb{R}$. It is also not difficult
to see from \eqref{inftythm2.6.2} and \eqref{vardefs3} that
\eqn{inftythm2.11}{
G = \frac{z^{4}}{2} \| \dz{U} \|^{2}+ z^{4} \langle \langle
\dz{U} | (n-3)z^{n-4}\phi + z^{n-3}\dz{\phi} \rangle \rangle
+ \frac{1}{2}\|(n-3)z^{n-4}\phi + z^{n-3}\dz{\phi}\|
}
and 
$P = P_{p} + z^{n-2}\sum_{j=1}^{4} P_{R,j}(\phi,X,m_{\infty},z)$
where $P_{R,j} : V_{2}\times E_{+} \times \mathbb{R} \times \mathbb{R}
\rightarrow \mathbb{R} \quad j=1,2,3,4$
are analytic maps that satisfy
$P_{R,j}(\epsilon Y_{1},Y_{2},x_{1},x_{2}) =
\epsilon^{j}P_{R,j}( Y_{1},Y_{2},x_{1},x_{2})$
for all $\epsilon \in \mathbb{R}$. Note also that 
$N = N_{p} -2z^{n-2}\sigma $.

Let
\leqn{inftythm2.16}{
 \omega := \dz{\phi} \AND \theta := z^{-1}\phi \: .
}
Using the above results, straightforward calculation shows that
there exists analytic maps
$\Gc : V_{2}\times V_{2} \times E_{+} \times \mathbb{R}^{3}
\rightarrow V_{2}$ and 
$\Mc :  V_{2}\times V_{2} \times E_{+} \times \mathbb{R}^{3}
\rightarrow \mathbb{R}$
such that equation \eqref{zfeq1} and \eqref{zfeq3} can be written as
\lalign{inftythm2.18}{
z\dz{\sigma} &= -(n-3)\sigma + z^{2}\Mc(\theta,\omega,X,m_{\infty},\sigma,z)
\: ,\label{inftythm2.18.1} \\
z\dz{\theta} &= -\theta + \omega \: ,\label{inftythm2.18.2}
}
and
\lalign{inftythm2.19}{
N\bigl(z\dz{\omega}+2(n-3)\omega & +(n-3)(n-4)\theta\bigr)
+2(n-3)\theta  \notag \\
& + 2\omega - A_{2}(\theta) - z\Gc(\theta,\omega,X,m_{\infty},\sigma,z) 
= 0\: . \label{inftythm2.19.2}
}
We can rewrite \eqref{inftythm2.19.2} as
\leqn{inftythm2.20}{
z\dz{\omega} = -2(n-2)\omega + \bigl(A_{2}-(n-3)(n-2)\id\bigr)\theta
+ z\hat{\Gc}(\theta,\omega,X,m_{\infty},\sigma,z)
}
where
\alin{inftythm2.21}{
\hat{\Gc}(Y_{1},Y_{2},Y_{3},x_{1},x_{2},x_{3})
 =  \frac{1}{x_{3}}&\left(\frac{1}{1-2(M(Y_{3},x_{1})+
x_{3}^{n-3}x_{2})x_{3}} - 1 \right)
\bigl(2(n-3)Y_{1}+ \bigr.\\
& \bigl. 2Y_{2}-A_{2}(Y_{1})\bigr)  + 
\Gc((Y_{1},Y_{2},Y_{3},x_{1},x_{2},x_{3}) \: .
}
Because $\Gc$ is analytic, it is clear that $\hat{\Gc}$ is
analytic in a neighborhood of 
$(0,0,X,m_{\infty},0,0) \in V_{2}\times V_{2} \times
E_{+}\times \mathbb{R}^{3}$.
For $Y\in V_{2}$ and $s\in \Ec$, define
$Y^{+}_{s} := \pr_{+}^{\tilde{s}} Y$ and
$Y^{0}_{s} := \pr_{0}^{\tilde{s}} Y$.
Recalling that $\pr_{0}^{\tilde{s}}A_{2} = A_{2} \pr_{0}^{\tilde{s}} = 0$ and
$\pr_{+}^{\tilde{s}}A_{2} = A_{2} \pr_{+}^{\tilde{s}} = 
s(s+1) \pr_{+}^{\tilde{s}}$
for every $s\in \Ec$, we can  write \eqref{inftythm2.18.2} and  
\eqref{inftythm2.20} as
\lalign{inftythm2.23}{
z\dz{\theta}{}^{+}_{s} &= -\theta^{+}_{s} + \omega^{+}_{s} \: ,
\label{inftythm2.23.1} \\
z\dz{\omega}{}^{+}_{s} &= -2(n-2)\omega^{+}_{s} + 
(s(s+1)-(n-3)(n-2))\theta^{+}_{s}
+ z\hat{\Gc}{+}_{s}(\theta,\omega,X,m_{\infty},\sigma,z) \:,
\label{inftythm2.23.2} \\
z\dz{\theta}{}^{0}_{s} &= -\theta^{0}_{s} + \omega^{0}_{s} \: ,
\label{inftythm2.23.3}  \\
\intertext{and}
z\dz{\omega}{}^{0}_{s} &= -2(n-2)\omega^{+}_{s} +
-(n-3)(n-2)\theta^{0}_{s}
+ z\hat{\Gc}^{0}_{s}(\theta,\omega,X,m_{\infty},\sigma,z) \: ,
\label{inftythm2.23.4}
}
for all $s\in \Ec$. 
For every $s\in\Ec$, introduce one last change of variables
\gath{inftythm2.24}{
\zeta^{1}_{s} := (-n+3)\theta^{0}_{s} - \omega^{0}_{s}\:,\quad
\zeta^{2}_{s} := (n-2)\theta^{0}_{s} + \omega^{0}_{s} \:, \\
\eta^{1}_{s} := \frac{1}{2s+1}((s-n+3)\theta^{+}_{s} - \omega^{+}_{s})
\:, \quad
\eta^{2}_{s} := \frac{1}{2s+1}((s+n-2)\theta^{+}_{s} + \omega^{+}_{s})
\: .
}
and let
$\zeta^{j} := \sum_{s\in\Ec} \zeta_{s}^{j}$ and
$\eta^{j} := \sum_{s\in\Ec} \eta_{s}^{j} \quad j=1,2$. 
Using this transformation we see from the above results that
(\ref{inftythm2.23.1}-\ref{inftythm2.23.4}) can be written as
\lalign{inftythm2.26}{
z\dz{\zeta}{}^{j}_{s} & = -(n-j)\zeta^{j}_{s} + z\Kc^{j}_{s}
(\zeta^{1},\zeta^{2},\eta^{1},\eta^{2},X,m_{\infty},\sigma,z) &&
\forall \; s\in\Ec \; , j=1,2 \; , \label{inftythm2.26.1} \\
z\dz{\eta}{}^{j}_{s} & = -(n-(-1)^{j}s-j)\eta^{j}_{s} + z\Hc^{j}_{s}
(\zeta^{1},\zeta^{2},\eta^{1},\eta^{2},X,m_{\infty},\sigma,z) &&
\forall \; s\in\Ec \; , j=1,2 \: . \label{inftythm2.26.2}
}
where $\Kc^{j}_{s}$ and $\Hc^{j}_{s}$ $(j=1,2)$ are $E_{0}$ and $E_{+}$
valued maps, respectively, that are analytic in a neighborhood
of $(0,0,X,m_{\infty},0,0) \in V_{2}\times V_{2} \times
E_{+}\times \mathbb{R}^{3}$. 
The system of differential equations given by 
\eqref{inftythm2.18.1}, \eqref{inftythm2.26.1}, and
\eqref{inftythm2.26.2} is equivalent to the original
system \eqref{zfeq1}, \eqref{zfeq3}. Moreover, if we
choose $n=\max\{3,3+\max\Ec\}$, then \eqref{inftythm2.18.1}, 
\eqref{inftythm2.26.1}, and \eqref{inftythm2.26.2}
are in a form to which  theorem \ref{BFM} applies. 
Applying this theorem shows that there exist a
unique solution $\{\sigma(z,a,Y),\zeta^{j}_{s}(z,a,Y),
\eta^{j}_{s}(z,a,Y)\}$ that is analytic
a neighborhood of $(z,a,Y) = (0,m_{\infty},X)$ and that satisfies
$\zeta^{j}_{s}(z) = \text{O}(z)$ , $\zeta^{j}_{s}(z) = \text{O}(z)$, 
and $\sigma(z) = \text{O}(z)$.
It then follows from  \eqref{inftythm2.5} that
$\pr^{\tilde{s}}_{+}(\Lp-\Op) = Y_{s} z^{s} + \text{O}(r^{s})$, 
$\pr^{\tilde{s}}_{0}(\Lp) = \text{O}(z^{s})$, and
$m  = a + \text{0}(z^{3})$.
\end{proof}

\begin{thm} \label{inftythm3} \mnote{[inftythm3]}
Every solution from theorem \ref{inftythm2} satisfies
equation \eqref{feq4} on a neighborhood of $r^{-1}=0$.
\end{thm}
\begin{proof}
Let $z=1/r$ and $\{\Lp(z),m(z)\}$ be a solution of the equations
\eqref{feq1} and \eqref{feq3} on a neighborhood $\Nc$ of $z=0$, which
we know exists by theorem \ref{inftythm2}. From lemma 
\ref{constraint} it is easy to see that in terms of the $z$ variable
\leqn{inftythm3.1}{
\gamma(z) = -z^{2}([\Lp(z),\dz{\Lambda}_{-}(z)] + [\Lm(z),\dz{\Lambda}_{+}(z)])
} 
and $\gamma(z)$ satisfies
$\dz{\gamma}(z) = f(z)\gamma(z)$
where $ f(z) = 2 N^{-1}(m-2zP)$.
By theorem \ref{inftythm2} and shrinking $\Nc$ if necessary, 
we see that $f(z)$ is analytic on $\Nc$. Therefore we can solve
$\dz{\gamma}(z) = f(z)\gamma(z)$ to get
$\gamma(z) = \gamma(0) \exp(\int_{0}^{z} f(\tau) d\tau)$ for all
$z \in \Nc$ .
But \eqref{inftythm3.1} shows that $\gamma(0)=0$. Therefore
$\gamma(z)=0$ for all $z\in \Nc$ and the theorem is proved. 
\end{proof}

\subsect{bh}{Regular black hole solutions}

\begin{thm} \label{bhthm1} \mnote{[bhthm1]}
Let $t=r-\rh$ and suppose $X\in V_{2}$ satisfies
\eqn{bhthm1.1}{
\nu(X) := \frac{1}{\rh}-\frac{1}{\rh}\norm{\Lo+[X,c(X)]}^{2} > 0 \: .
}

Then there exist a unique solution
$\{\Lp(t,Y), N(t,Y)\}$ to the system of differential
equations \eqref{feq1} and \eqref{feq3} that is
analytic in a neighborhood of $(0,X)$
in $\mathbb{R}\times V_{2}$ and satisfies
\eqn{bhthm1.2}{
N(t) = \nu(Y) t + \text{0}(t^{2})  \AND
\Lp(t) = Y + \text{0}(t) \: .
}
\end{thm}

\begin{proof}
The proof is exactly the same as the proof of theorem 6 of \cite{eymg} with
$E_{+}$ replaced by $V_{2}$.
\end{proof}

\begin{thm} \label{bhthm2} \mnote{[bhthm2]}
Every solution from theorem \ref{bhthm1} satisfies
equation \eqref{feq3} on a neighborhood of $r=\rh$.
\end{thm}
\begin{proof}
Let $t=r-\rh$ and suppose $X\in V_{2}$ satisfies
$\nu := {\rh}^{-1}-{\rh}^{-1}\norm{\Lo+[X,c(X)]}^{2} > 0$ .
Then we know by the previous theorem that there exists a solution $\{\Lp(t),
N(t)\}$ to the system of differential equations \eqref{feq1} and \eqref{feq3}
that is analytic in a neighborhood of $t=0$ and satisfies
\leqn{bhthm2.2}{
N(t) = \nu t + \text{0}(t^{2})  \AND
\Lp(t) = X + \text{0}(t) \: .
}
Let
$f(t) := -2t\left((t+\rh)^{2}N(t)\right)^{-1}\left(m(t)-
(t+\rh)^{-1}P(t)\right)$.
Then \eqref{bhthm2.2} shows that $f(t)$ is analytic in a neighborhood of
$t=0$.  A short calculation shows that $f(0)=-1$, and therefore we can write
$f(t) = -1 + tg(t)$
where $g(t)$ is analytic in near $t=0$. 
Consider the differential equation on $V_{2}$,
\leqn{bhthm2.5}{
t\frac{d\eta(t)}{dt} = -\eta(t) + tg(t)\eta(t) \: .
}
It has $\eta(t)=0$ as a solution, and therefore this is the unique analytic
solution near $t=0$ by theorem \ref{BFM}.  But lemma \ref{constraint} shows
that $\gamma(t) = [\Lp(t),\frac{d\Lm}{dt}(t)] + [\Lm(t),\frac{d\Lp}{dt}(t)]$
also solves \eqref{bhthm2.5} in an neighborhood of $t=0$.  Because $\gamma(t)$
is analytic near $t=0$, we must have $\gamma(t)=0$ near $t=0$.
\end{proof}

\sect{examples}{A numerical example}

So far we have not yet found a global numerical solution that has the
correct fall off behavior for $r\ra\infty$.  The following example is
for the gauge group $SO(5)$ ($B_{2}$) for the action with
characteristic $(20)$, one of the simplest irregular cases that will
not reduce. Figure 1 %\ref{B2:20cen}
shows a solution near the center with real initial values $w_{\alpha}(0)$ 
which develops nonzero imaginary components $v_{\alpha}$. Figure 2 %\ref{B2:20inf}
shows a solution for large $r$. As is apparent the function $N$ decreases 
from its value $1$ at infinity as $r$ decreases down to some minimum, but
then increases rapidly. Only by very careful tuning of the data at
infinity one can possibly avoid this behavior and construct a
`physical' solution in which $N$ alwas remains between 0 and 1. For a
globally bounded solution we have also indications that necessarily
$\Vert\Lp\Vert \leq \Vert\Op\Vert$.

\begin{center}
\begin{figure}\label{B2:20cen}
\epsfig{file=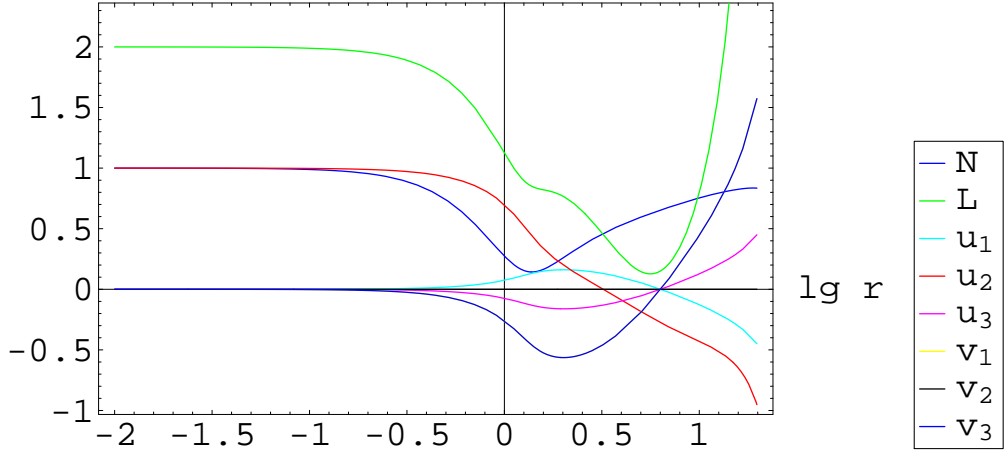,height=3.in,width=5.5in}
\caption{Gauge group $SO(5)$ or $B_{2}$: characteristic $(20)$,
$\Lo=2\hb_1+\hb_2$, $(w_{\alpha}(0))=(0,1,0)$, $\Vert\Op\Vert^{2}=2$,
$\beta=(0.1,-0.35,-0.28)$. The function $v_{2}$ is identically 0,
$u_3=-u_1$, and $v_3=-v_1$. The quantity L is $\Vert\Lp\Vert^{2}$. 
For a globally regular solution it should nowhere exceed its value at the
center and at infinity.}
\end{figure}
\end{center}

\begin{center}
\begin{figure}\label{B2:20inf}
\epsfig{file=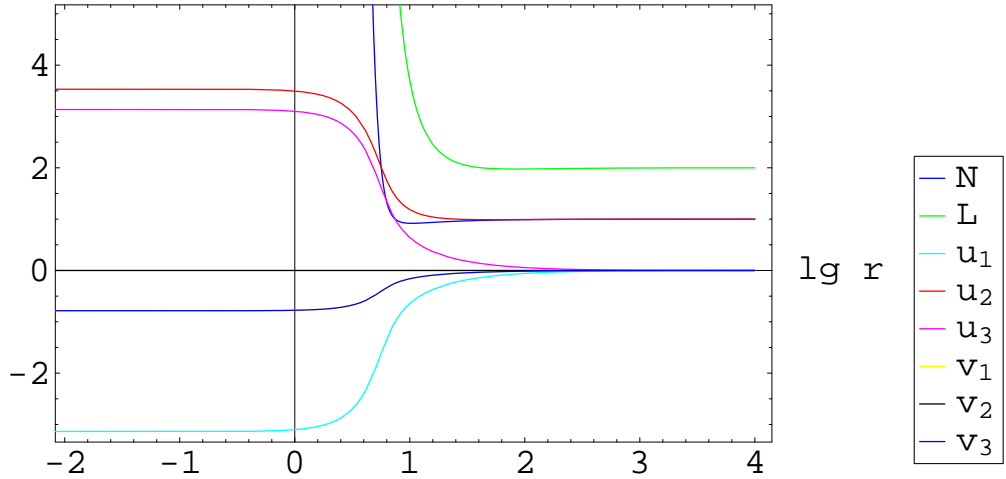,height=3.in,width=5.5in}
\caption{Gauge group $SO(5)$ or $B_{2}$: characteristic $(20)$,
$\Lo=2\hb_1+\hb_2$, $(w_{\alpha}(\infty))=(0,1,0)$,
$\Vert\Op\Vert^{2}=2$, $m_{\infty}=0.52$, $\alpha=(-8,-2,-1)$. The
function $v_{2}$ is identically 0, $u_3=-u_1$, and $v_3=-v_1$. The
quantity L is $\Vert\Lp\Vert^{2}$. For a globally regular solution it
should nowhere exceed its value at the center and at infinity. Also
$N$ should remain between 0 and 1.}
\end{figure}
\end{center}

\end{document}